\documentclass[%
 reprint, $preprint, reprint, or letter$
 amsmath,amssymb,
aps,
longbibliography,
]{revtex4-2}

\usepackage{graphicx}
\usepackage{dcolumn}
\usepackage{physics}
\usepackage{gensymb}
\usepackage{bm}
\usepackage{xcolor}
\usepackage{footmisc}
\usepackage{xr}

\begin{document}

\preprint{APS/123-QED}

\title{Fast relaxation on qutrit transitions of nitrogen-vacancy centers in nanodiamonds}

\author{A.~Gardill}
\thanks{These authors contributed equally.}

\author{M.~C.~Cambria}
\thanks{These authors contributed equally.}

\author{S.~Kolkowitz}
\email{kolkowitz@wisc.edu}
 
\affiliation{%
 Department of Physics, University of Wisconsin, Madison, Wisconsin 53706, USA
}%

\date{\today}
\begin{abstract}
Thanks to their versatility, nitrogen-vacancy (NV) centers in nanodiamonds have been widely adopted as nanoscale sensors. However, their sensitivities are limited by their short coherence times relative to NVs in bulk diamond. A more complete understanding of the origins of decoherence in nanodiamonds is critical to improving their performance. Here we present measurements of fast spin relaxation on qutrit transitions between the energy eigenstates composed of the $m_s = \ket{\pm1}$ states of the NV$^-$ electronic ground state in $\sim40$-nm nanodiamonds under ambient conditions. For frequency splittings between these states of $\sim20$~MHz or less the maximum theoretically achievable coherence time of the NV spin is $\sim2$~orders of magnitude shorter than would be expected if the NV spin is treated as a qubit. 
We attribute this fast relaxation to electric field noise. We observe a strong falloff of the qutrit relaxation rate with the splitting between the states, suggesting that, whenever possible, measurements with NVs in nanodiamonds should be performed at moderate axial magnetic fields ($>60$ G). We also observe that the qutrit relaxation rate changes with time. These findings indicate that surface electric field noise is a major source of decoherence for NVs in nanodiamonds.


\end{abstract}

\maketitle

\section{Introduction}

Nitrogen-vacancy (NV) centers in diamond have a number of properties that make them attractive candidates for use as quantum sensors. The NV$^-$ ground-state electronic spin triplet can be optically polarized and readout via fluorescence, exhibits millisecond-long coherence times at room temperature in bulk diamond \cite{balasubramanian2009, deLange2010universal, Naydenov2011, bar_gill2013}, and can be used as a probe of the local magnetic \cite{Rondin2014}, electric \cite{Dolde2011, Mittiga2018}, strain \cite{Ovartchaiyapong2014,Teissier2014}, and thermal \cite{Kucsko2013, Toyli2013} environments. NVs in diamond nanocrystals, or nanodiamonds, are particularly attractive for sensing applications as they can be functionalized \cite{Kruger2006}, placed at the ends of scanning tips \cite{Balasubramanian2008, Rondin2014}, deterministically positioned on nanophotonic structures \cite{Bogdanov2019, Schietinger2009}, optically levitated \cite{Neukirch2013,Hoang2016}, or even inserted into living cells \cite{Kucsko2013}. Unfortunately, the coherence times of the electronic spins of NVs in commercial nanodiamonds are consistently on the order of $\sim1-10$ microseconds \cite{Naydenov2011, Song2014}, $\sim2-3$ orders of magnitude shorter than what is regularly achieved with NVs in bulk diamond \cite{balasubramanian2009, deLange2010universal, Naydenov2011, bar_gill2013}, limiting their sensitivities for many applications. There have been a number of efforts to improve the coherence times of NVs in nanodiamonds, including milling from high purity bulk diamond \cite{Knowles2014}, high temperature annealing \cite{tsukahara2019}, dynamical decoupling \cite{Knowles2014, Liu2014}, and a variety of surface treatments \cite{Song2014, Brandenburg2018, Ryan2018}, but thus far only marginal improvements have been observed. A better understanding of the origins of decoherence in nanodiamonds is required to unlock their full potential.

\begin{figure}[b]
\vspace{-0.3cm}
\includegraphics{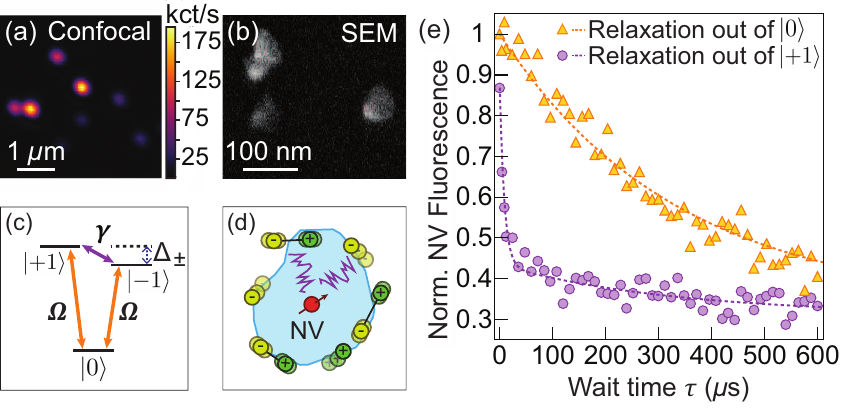}
\caption{\label{fig:one} Fast relaxation on qutrit spin transitions of NVs in nanodiamonds. Note that in the figure \(\ket{m_s}\) refers to state \(\ket{H; m_s}\). (a) Confocal microscope image of NV centers in nanodiamonds. (b) Scanning electron microscope image of three nanodiamonds. (c) Ground state level-structure of the NV$^-$. Qubit transitions between \(\ket{H;0} \leftrightarrow \ket{H;\pm1}\) occur at rate \(\Omega\), and qutrit transitions between \(\ket{H;+1} \leftrightarrow \ket{H;-1}\) at rate \(\gamma\). (d) Diagram of a single NV in a nanodiamond. Moving charges or fluctuating electric dipoles create electric field noise (purple lines) that drive NV spin relaxation. (e) Measurement of state dependent population relaxation. Relaxation out of \(\ket{H;0}\) (orange triangles) exhibits a single exponential with rate \(3\Omega\). Relaxation out of \(\ket{H;+1}\) (purple circles) exhibits a biexponential decay that depends on both \(\Omega\) and \(\gamma\) (dashed purple line). Dashed lines are Eqs.~\ref{population_zero},~\ref{population_plus}, and account for $\pi$-pulse infidelities, with \(\Omega\) = 1.0 kHz, \(\gamma\) = 56 kHz.}
\end{figure}

Prior efforts to improve the coherence times of NVs in nanodiamonds have focused on magnetic noise as the primary source of both dephasing and relaxation \cite{Knowles2014,Song2014, Liu2014, Brandenburg2018, Ryan2018}. However, in a recent work Myers \textit{et al.} found that in shallow NVs $\sim7$ nm away from the surface in bulk diamond, electric field noise emanating from the surface can drive magnetic-dipole-forbidden transitions with $\Delta m_s = 2$ between the $m_s = \ket{\pm1}$ states, sometimes called double quantum transitions, at rates up to $\sim2$~kHz, or more than $20\times$ the rate between the $m_s=\ket{0}$ state and the $m_s = \ket{\pm1}$ states \cite{Myers2017}. As NVs in nanodiamonds are tens of nanometers away from the surface in all directions and the surfaces are more difficult to control than those of bulk diamond samples, it is natural to ask whether this effect occurs in nanodiamonds as well. 

Because nanodiamonds frequently exhibit significant intrinsic d.c.~strain or electric fields, it is especially important in this context to distinguish between the eigenstates of \(S_{z}\) and the eigenstates of the NV spin Hamiltonian. We use \(\ket{S_{z};m_{s}}\) to denote the eigenstate of \(S_{z}\) with spin projection \(m_{s}\). Similarly, we use \(\ket{H;m_{s}}\) to denote the eigenstate of the Hamiltonian with majority component \(\ket{S_{z};m_{s}}\). For simplicity, in Figs.~\ref{fig:one} and \ref{fig:two} we refer to the state \(\ket{H;m_{s}}\) as \(\ket{m_{s}}\). We will refer to the \(\ket{H;-1} \leftrightarrow \ket{H;+1}\) transition as the qutrit transition, a generalization of the \(\ket{S_{z};-1} \leftrightarrow \ket{S_{z};+1}\) double quantum transition \cite{Myers2017}. Likewise we will refer to the \(\ket{H;0} \leftrightarrow \ket{H;\pm1}\) transitions as the qubit transitions.

In this paper we present measurements indicating that relaxation on the qutrit transition is a dominant source of decoherence for NVs in $\sim40$-nm commercial nanodiamonds under ambient environmental conditions. At low axial magnetic fields, $B_z<10$~G, we find that the rate of relaxation on the qutrit transition can exceed 100~kHz, more than two orders of magnitude faster than the rate on the qubit transitions, limiting the maximum theoretically achievable coherence times of NVs in this regime to tens of microseconds. We attribute this to electric field noise that drives the transitions between the $\ket{H;\pm1}$ states. We observe this behavior in all 5 of the single NVs in nanodiamonds that we studied. We characterize the dependence of this rate on the frequency splitting, $\Delta_{\pm}$, between the $\ket{H;\pm1}$ states and observe a strong falloff with  $\Delta_{\pm}$, consistent with a $1/f^2$ scaling of the noise power spectral density. This indicates that, whenever possible, coherent measurements with nanodiamonds should be performed with moderate magnetic fields ($>60$ G) applied along the NV axis. Finally, we observe fluctuations in this relaxation rate on hour to day time-scales, which we consider to be a strong indication that the noise is emanating from the nanodiamond surface.

\section{Experimental Methods}

\subsection{Samples and apparatus}
 
All data was taken with commercial nanodiamonds from Ad\'{a}mas Nano, which are crushed from high-pressure high-temperature monocrystalline microdiamonds \footnote{See the product details from Ad\'{a}mas Nano for more information about these nanodiamonds with item number NDNV40nmLw10ml}. Ad\'{a}mas reports that the nanodiamonds have a substitutional nitrogen concentration of $\sim$~100 ppm before electron irradiation and $\sim$~60-80 ppm after irradiation and annealing. Ad\'{a}mas annealed the nanodiamonds in vacuum at $850^\circ$C and then oxidized the graphitic layer in acids before suspending the nanodiamonds in solution. The nanodiamond surface is expected to be carboxylated. After receiving the nanodiamond solution, no additional treatment was performed on the nanodiamond surface. Our results are therefore of particular relevance to researchers using similar commercial nanodiamonds without further sample processing. The nanodiamonds have a mean diameter of 40 nm and the diameter distribution has a full width at half maximum of $\sim$~40 nm. Each nanocrystal contains an average of 1 - 4 NVs. 

The nanodiamonds were diluted in deionized (DI) water to a concentration of 10 $\mu$g/mL. To increase the solution's adhesion to the substrate, poly-vinyl alcohol (PVA) was added with a weight-to-weight (PVA to DI water) concentration of 0.17\%. A gridded glass coverslip was cleaned with isopropyl alcohol and nanodiamond solution was spin coated onto the coverslip at 3000 rpm for 20 seconds. The NVs were imaged through the back side of the coverslip, with the nanodiamonds in air under ambient conditions.

 Our experimental setup consists of a room temperature confocal microscope. A fluorescence image of single NVs in nanodiamonds is shown in Fig.~\ref{fig:one}(a), along with a scanning electron microscope image of nanodiamonds from the same suspension in Fig.~\ref{fig:one}(b). Before measuring relaxation rates, we select nanodiamonds containing only single photon emitters determined by second-order photon correlation measurements \footnote{See supplemental material for experimental details and additional data.}. We then perform optically detected magnetic resonance (ODMR) to select for NVs. A majority of the emitters showed low or no contrast ODMR signal. Those NVs with a measurable ODMR signal at low magnetic fields were used for experiments (see Table~\ref{tab:rel_rates}). In total, we identified 5 single NVs with measurable signals out of a starting set of 110 nanodiamonds.

\begin{figure}[b]
\includegraphics{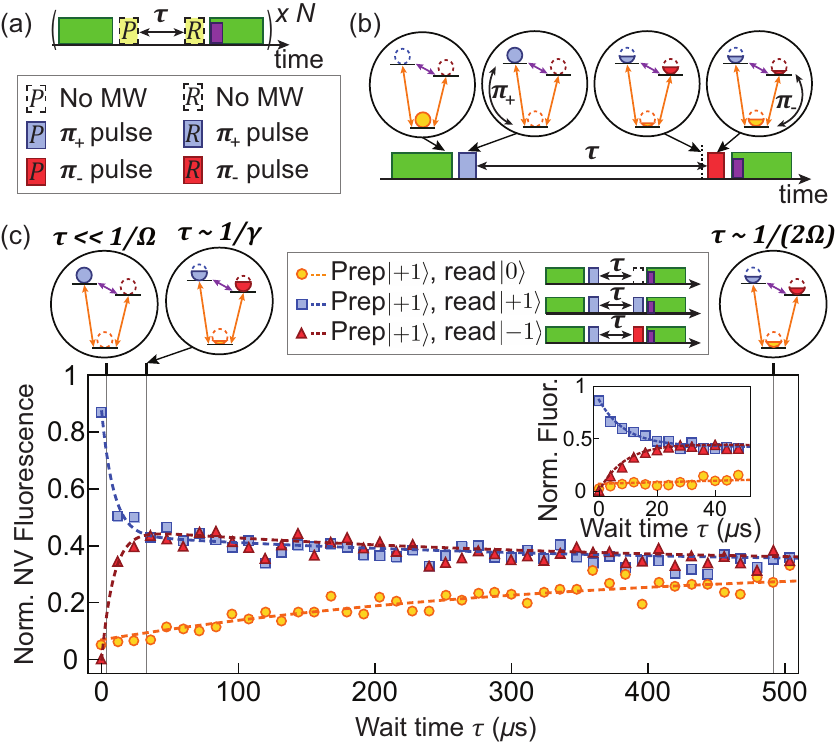}
\caption{\label{fig:two} State-selective population relaxation measurements. Note that in the figure \(\ket{m_s}\) refers to state \(\ket{H; m_s}\). (a) The measurement sequence consists of preparing NV in \(\ket{H;\pm1}\) or \(\ket{H;0}\), waiting time \(\tau\), and reading out the population of \(\ket{H;\pm1}\) or \(\ket{H;0}\). Sequences are repeated \(N \sim\)~10\(^5\) times. Preparation and readout of population in \(\ket{H;0}\)  is achieved using 532~nm optical illumination (green rectangles) and fluorescence readout (purple rectangle), while state-selective microwave \(\pi\)-pulses transfer population into and out of \(\ket{H;\pm1}\) (red and blue rectangles). (b) Example of single measurement sequence of preparation in \(\ket{H;+1}\)  and readout in \(\ket{H;-1}\). 
(c) Representative population dynamics for a single NV in a nanodiamond prepared in \(\ket{H;+1}\), with $\Delta_{\pm} = 28.9(6)$~MHz. The population in states \(\ket{H;+1}\) (blue squares), \(\ket{H;-1}\) (red triangles), and \(\ket{H;0}\) (yellow circles) after wait time \(\tau\) are measured by the method explained in (a) and (b). Dashed colored lines are solutions to the three level population dynamics equations, accounting for \(\pi\)-pulse infidelities, with \(\Omega\) = 1.0 kHz, \(\gamma\) = 56 kHz. Inset shows population dynamics of same NV over first 50 \(\mu\)s of relaxation.}
\end{figure}

\begin{table*}
\caption{\label{tab:rel_rates}
Characteristics of the five measured NVs. For each NV, the maximum \(\gamma\) we observed is reported with the corresponding splitting \(\Delta_{\pm}^{\gamma}\), the resulting \(T_\text{2,max}\), and the contrast of the pulsed ODMR resonance at the splitting \(\Delta_{\pm}^{\gamma}\). The calculated average \(\Omega_\text{avg}\), \(\gamma_\infty\), and \(A_0\) for each NV are also listed (see Fig.~\ref{fig:three}), along with the measured splitting  in the absence of any applied magnetic field, \(\Delta_{\pm}^\text{zf}\). Reported error is twice the standard error.}

\begin{ruledtabular}
\begin{tabular}{c|cccc|cccc}
 & Max \(\gamma\) (kHz) & \(\Delta_{\pm}^{\gamma}\) (MHz) & \(T_\text{2, max}\) (\(\mu\)s) & ODMR Contrast & \(\Omega_\text{avg}\) (kHz) & \(\gamma_\infty\) (kHz) &\(A_0\) (MHz\(^2\cdot\)kHz) & \(\Delta_{\pm}^\text{zf}\) (MHz)\\ 
\hline
NV1 &  117(8) & 19.8(6) &16.6(11)& 36\% & 1.1(3) & 0.74(6) &  33.9(6)\(\times10^3\) & 19.5(6) \\

NV2 &  124(6) &15.3(6) & 16.0(8) & 40\% & 0.32(13) & 0.20(3) & 15.7(2)\(\times10^3\) & 15.3(6) \\

NV3 &  110(20) &17.1(6) & 18(3) & 20\% & 1.0(10) & 4.7(4) & 59(3)\(\times10^3\)  & 15.4(6)\\

NV4 &  35(3) &23.4(6) & 57(4) & 31\% & 0.30(18) & 0.56(4) & 29.4(6)\(\times10^3\) & 23.4(6) \\

NV5 &  240(50) &10.9(6) & 8.3(17) & 10\% & 0.7(6) & 3.7(5) & 18.7(18)\(\times10^3\) & 6.9(6) \\

\end{tabular}
\end{ruledtabular}
\end{table*}

\subsection{Measurement sequence}

As shown in Fig.~\ref{fig:one}(c), the NV electronic ground state is a spin-triplet. In most NV studies and applications, a d.c. magnetic field ($B_z$) is applied along the NV axis to lift the degeneracy between the $\ket{H;\pm1}$ states, and the NV is then treated as a two-level system with a spin lifetime $T_1$ found by measuring the lifetime of the $\ket{H;0}$ state. Figure~\ref{fig:one}(e) shows representative population decay curves for the spin states $\ket{H;0}$ and $\ket{H;+1}$ of an NV in a nanodiamond. At low $B_z$, population prepared and readout in $\ket{H;+1}$ exhibits a biexponential decay and has mostly depolarized after just $\sim20$ microseconds (purple circles). However, the standard $T_1$ measurement employed in almost all NV studies, which consists of optically polarizing in $\ket{H;0}$, waiting some time $\tau$, and measuring the population in $\ket{H;0}$ via fluorescence, exhibits a single exponential decay with a much longer time-constant of $\sim330$ microseconds (orange triangles). Critically, this measurement is blind to the population leakage between $\ket{H;+1}$ and $\ket{H;-1}$, and would therefore drastically overestimate the achievable coherence time $T_2$ for this NV.

Figure~\ref{fig:two}(a) illustrates the measurement sequence we use to measure the population dynamics into and out of each of the three spin states \cite{Kolkowitz2015,Myers2017,Sangtawesin2019}. Preparation and readout of any of the three states is achieved by using a combination of 532 nm optical polarization and state-selective resonant microwave $\pi$-pulses, allowing for a total of nine measurement combinations \footnotemark[2]. For example, Figure~\ref{fig:two}(b) shows the sequence used to prepare in $\ket{H;+1}$ and measure the population in $\ket{H;-1}$. After the population is optically polarized into $\ket{H;0}$, a microwave $\pi_{+}$-pulse transfers the population to $\ket{H;+1}$. After a wait time $\tau$, a $\pi_{-}$-pulse swaps the populations in $\ket{H;-1}$ and $\ket{H;0}$. The transferred population of the state $\ket{H;-1}$ is then readout out via fluorescence under 532 nm illumination and normalized to a reference measurement of the NV brightness when it has been polarized in the $\ket{H;0}$ state. The purple circles in Fig.~\ref{fig:one}(e) correspond to using this sequence to prepare in $\ket{H;+1}$, wait, and then readout in $\ket{H;+1}$. 

Figure~\ref{fig:two}(c) shows a measurement of the population dynamics of the three spin states after preparation in $\ket{H;+1}$ for a representative NV at a splitting of $\Delta_{\pm}~=~28.9(6)$ MHz. The fast relaxation out of $\ket{H;+1}$ into $\ket{H;-1}$ is readily apparent. Remarkably, the population in $\ket{H;-1}$ is non-monotonic with time, as the population prepared in $\ket{H;+1}$ first rapidly decays to an even mixture of $\ket{H;\pm1}$ before slowly decaying into an unpolarized mixture of all three spin states.

\subsection{Isolation of rates $\gamma$ and $\Omega$}

To capture the full population dynamics, we solve the rate equations for a generic three-level system with arbitrary transition rates between each pair of states. Empirically, we find that the transition rates for $\ket{H;0} \leftrightarrow \ket{H;+1}$ and $\ket{H;0} \leftrightarrow \ket{H;-1}$  are equivalent to within our measurement uncertainties \footnotemark[2], which simplifies the analysis. Defining the qutrit transition rate as $\gamma$ and the qubit transition rate as $\Omega$ (see Fig.~\ref{fig:one}(c)), the population equations for the three NV states are given by \cite{Kolkowitz2015, Myers2017, Ariyaratne2018}
\begin{equation}\label{population_zero}
\rho_0(\tau) = \frac{1}{3} + \left(\rho_0(0) - \frac{1}{3}\right) e^{-3\Omega \tau},
\end{equation}
\begin{align}\label{population_plus}
\rho_{\pm1}(\tau) &= \frac{1}{3} \pm \frac{1}{2}\Delta\rho_{\pm1}(0) e^{-(2\gamma +\Omega) \tau} \\ 
& \quad \quad - \frac{1}{2} \left(\rho_0(0) - \frac{1}{3}\right) e^{-3\Omega \tau}\nonumber,
\end{align}
where $\tau$ is the wait time, $\rho_{0,\pm1}(\tau)$ are the populations of the $\ket{H;0,\pm1}$ states at time $\tau$, $\rho_0(0)$ is the initial population of $\ket{H;0}$, and $\Delta\rho_{\pm1}(0)$ is the difference in the initial population of $\ket{H;+1}$ and $\ket{H;-1}$. The dashed colored lines in Fig.~\ref{fig:one}(e) and~\ref{fig:two}(c)  are plots of Eqs.~\ref{population_zero} and \ref{population_plus}, accounting for measured $\pi$-pulse infidelities \footnotemark[2]. We observe excellent agreement between this model and all of the relaxation measurements we performed.


To extract values for the transition rates $\gamma$ and $\Omega$, we follow the analysis protocol laid out in \cite{Myers2017}. We denote the time-dependent populations we measure in state $\ket{H;j}$ after initial preparation in $\ket{H;i}$ with $P_{i,j}(\tau)$. From Eqs.~\ref{population_zero} and \ref{population_plus}, subtracting $P_{0,0}(\tau)$ and $P_{0,+1}(\tau)$ gives 
\begin{equation}\label{omega_subtracted_eq}
F_\Omega(\tau) = P_{0,0}(\tau) - P_{0,+1}(\tau) = a e^{-3\Omega\tau},
\end{equation}
\noindent where $a$ is the fluorescence contrast between the two subtracted data sets. Similarly, subtracting $P_{+1,+1}(\tau)$ and $P_{+1,-1}(\tau)$ results in 
\begin{equation}\label{gamma_subtracted_eq}
F_\gamma(\tau) =  P_{+1,+1}(\tau) - P_{+1,-1}(\tau) = a e^{-(2\gamma + \Omega)\tau}.
\end{equation}
\noindent This method allows us to fit a single exponential to each subtracted set and isolate the rates $\Omega$ and $\gamma$. For all 5 NVs we observe $\gamma\gg\Omega$ at low axial magnetic fields. Characteristics of each NV are summarized in Table~\ref{tab:rel_rates}.

\section{Results}

\begin{figure}[b]
\includegraphics{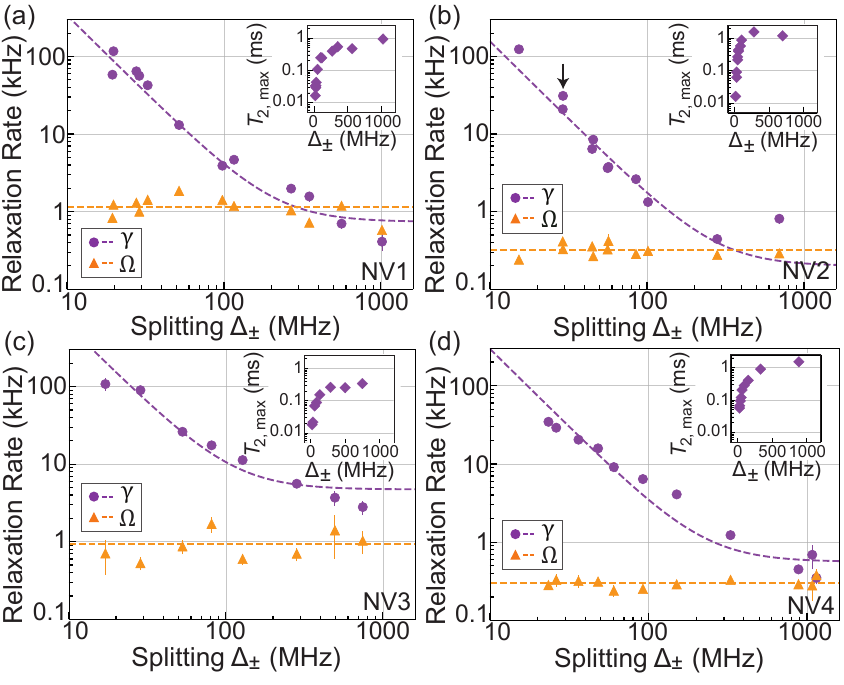}
\caption{\label{fig:three} Dependence of relaxation rates $\gamma$ and $\Omega$ on the frequency splitting between \(\ket{H;\pm1}\) states, \(\Delta_{\pm}\), for four separate NVs. Purple circles represent measurement of \(\gamma\), orange triangles represent measurement of \(\Omega\). A fit to \(\gamma(\Delta_{\pm})~=~A_0/\Delta_{\pm}^{2}~+~\gamma_\infty\) is shown on all four plots (dashed purple lines). Error bars are twice the standard error. The point in (b) marked by a black arrow is referenced in Fig.~\ref{fig:four}. Insets show the corresponding maximum theoretically achievable \(T_\text{2,max}\) (purple diamonds) based on Eq.~\ref{t2_max} on a semi-log plot.} 
\end{figure}

Figure~\ref{fig:three} shows the relaxation rates $\gamma$ and $\Omega$ as a function of the frequency splitting $\Delta_{\pm}$ for 4 of the NVs measured (similar rates and scalings were also observed for the 5th NV, see \footnotemark[2]). The rate $\gamma$ initially decreases rapidly with increasing $\Delta_{\pm}$ while $\Omega$ appears constant over the measured range $B_z \sim 0 - 200$~G. Similar to the results for shallow NVs in bulk diamond \cite{Myers2017}, the scaling is relatively well described by $\gamma(\Delta_{\pm})=A_0/\Delta_{\pm}^{2} + \gamma_\infty$ (dashed purple lines), where $A_0$ and $\gamma_\infty$ are constants. 

All 5 of the NVs studied here are sampled from a single spin-coated glass coverslip. To verify that the noise we observe is not a result of the glass surface or of the PVA in the solution, we measure a single NV deposited on a clean silicon wafer from a DI water/nanodiamond solution containing no PVA. The solution was dropped onto a clean silicon wafer and then heated on a hot-plate at $160^\circ$C to evaporate the water. We measure the relaxation rates $\gamma$~= 63(10)~kHz and $\Omega$ = 0.17(3)~kHz of this NV at splitting $\Delta_{\pm} = 13.9(6)~\text{MHz}$ \footnotemark[2]. These rates are similar to the rates measured on the other NVs at similar low splittings, confirming the fast relaxation behavior we observe is intrinsic to the nanodiamonds. Additionally, all measurements are taken with the same power of the 532 nm laser for state polarization and readout. To determine whether this intensity has any effect on the observed rates, we measure $\gamma$ and $\Omega$ of NV2 at a low splitting at four different laser powers. We find that the rates do not scale with the laser intensity \footnotemark[2].
 

\section{Discussion}

\subsection{NV Hamiltonian in the nanodiamond context}

To discuss the origins of the observed fast transitions between $\ket{H;+1}$ and $\ket{H;-1}$, we introduce the NV ground-state Hamiltonian. Ignoring hyperfine coupling, the Hamiltonian \cite{Doherty2012} can be expressed as a sum of the zero-field splitting $D_{gs}$, the interaction with magnetic field \(\boldsymbol{B}\), and the interaction with electric and scaled strain field \(\boldsymbol{\Pi}\). That is (with $\hbar=1$),
\begin{equation}\label{hamiltonian}
    H = H_{zfs} + H_{B} + H_{\Pi},
\end{equation}
\noindent where
\begin{align}
    H_{zfs} &= D_{gs}S_{z}^{2}, \label{hamiltonian_zfs}\\
    H_{B} &= g\mu_{B}\boldsymbol{B} \cdot \boldsymbol{S}, \label{hamiltonian_B}\\
    H_{\Pi} &= d_{\parallel} \Pi_{z}S_{z}^{2} + d_{\perp}'\Pi_{x}\left(S_{x}S_{z}+S_{z}S_{x}\right) \label{hamiltonian_Pi}\\
    &\quad+ d_{\perp}'\Pi_{y}\left(S_{y}S_{z}+S_{z}S_{y}\right) + d_{\perp}\Pi_{x}\left(S_{y}^{2}-S_{x}^{2}\right) \nonumber\\
    &\quad+ d_{\perp}\Pi_{y}\left(S_{x}S_{y}+S_{y}S_{x}\right) \nonumber.
\end{align}
Here, \(\boldsymbol{S}\) is the vector of spin-1 operators, \(D_{gs}/2\pi=2.87 \ \text{GHz}\) is the zero-field splitting between \(\ket{H;0}\) and \(\ket{H;\pm1}\), \(g\mu_{B}/2\pi=2.8 \ \text{MHz/G}\) is the NV electronic spin gyromagnetic ratio, and \(d_{\parallel}\), \(d_{\perp}\), and \(d_{\perp}'\) are electric dipole parameters with measured values \(d_{\parallel}/2\pi=0.35 \text{ Hz cm/V}\) and \(d_{\perp}/2\pi=17 \text{ Hz cm/V}\) \cite{vanoort1990}. To our knowledge, \(d_{\perp}'\) has not been measured, but \textit{ab initio} studies suggest \(d_{\perp}'\) and \(d_{\perp}\) may have similar values \cite{Doherty2012}. However, because the \(d_{\perp}'\) terms mix states with energy splitting \(\sim D_{gs}\), for weak electric fields \(d_{\perp}'\) is frequently taken to be zero \cite{vanoort1990, Doherty2012}. 

In the absence of strain, electric fields, and off-axis magnetic fields, the eigenstates of the Hamiltonian are the same as those of \(S_{z}\): $\ket{H;m_{s}} = \ket{S_{z};m_{s}}$. This identification has been made in previous studies of the qutrit transition in shallow NVs in bulk diamond \cite{Myers2017, Sangtawesin2019}. In this case, \(\bra{S_{z};-1}H_{B}\ket{S_{z};+1}=0\), indicating that the \(\ket{S_{z};-1} \leftrightarrow \ket{S_{z};+1}\) qutrit transition is magnetic-dipole-forbidden, while \(\bra{S_{z};-1}H_{\Pi}\ket{S_{z};+1}\) is nonzero. Relaxation on the \(\ket{S_{z};-1} \leftrightarrow \ket{S_{z};+1}\) transition can therefore be unambiguously attributed to resonant electric field/strain noise. Because nanodiamonds often contain large intrinsic d.c. strain or electric fields, the qutrit transition is, in general, not magnetic-dipole-forbidden. Specifically, at zero applied magnetic field the energy eigenstates of NVs affected by off-axis strain or electric fields are the bright and dark states $\ket{H;\pm}=\left(\ket{S_{z};+1}\mp e^{-i\phi_{\Pi_\perp}}\ket{S_{z};-1}\right)/\sqrt{2}$, where $\phi_{\Pi_\perp}$ is the angle of the strain/electric field in the plane transverse to the NV axis \cite{Mittiga2018}. The \(\ket{H;-} \leftrightarrow \ket{H;+}\) transition, like the \(\ket{S_{z};-1} \leftrightarrow \ket{S_{z};+1}\) transition, can be driven by electric field/strain noise. Calculation of the \(\bra{H;-}H_{B}\ket{H;+}\) matrix element demonstrates that on-axis magnetic noise can also drive the \(\ket{H;-} \leftrightarrow \ket{H;+}\) qutrit transition, allowing for the possibility that the fast relaxation rates observed at low applied magnetic field are due to transitions between mixed states driven by magnetic noise. We next investigate this possibility and conclude that it is unlikely.

\subsection{Effect of magnetic field noise on mixed eigenstates}

If we treat magnetic field noise as a harmonic perturbation and account for the anisotropy in the effect of the noise by averaging over the possible noise field orientations, then \(\gamma_{B}\), the first-order \(\ket{H;-1} \leftrightarrow \ket{H;+1}\) relaxation rate due to magnetic field noise, obeys 
\begin{equation}\label{gamma_B}
    \gamma_{B} \propto \overline{\abs{\bra{H;-1}H_{B}\ket{H;+1}}^{2}},
\end{equation}
where \(H_{B}\) is the perturbative magnetic Hamiltonian (Eq. \ref{hamiltonian_B}) and \(\overline{\ \cdot\ }\) denotes the average of all possible orientations. Similarly, \(\Omega_{B,\pm}\), the first-order \(\ket{H;0} \leftrightarrow \ket{H;\pm1}\) relaxation rate due to magnetic noise obeys
\begin{equation}\label{Omega_B}
    \Omega_{B,\pm} \propto \overline{\abs{\bra{H;0}H_{B}\ket{H;\pm1}}^{2}}.
\end{equation}


In order to determine whether the observed \(\ket{H;-1} \leftrightarrow \ket{H;+1}\) relaxation is attributable to magnetic field noise or electric field noise, we compare the ratio of \(\Omega\) and \(\gamma\) where \(\Delta_{0-}\), the splitting between \(\ket{H;0}\) and \(\ket{H;-1}\), is on the same order of magnitude as \(\Delta_{\pm}\), the splitting between \(\ket{H;-1}\) and \(\ket{H;+1}\). In this case, we expect that the magnetic noise power at frequency \(\Delta_{\pm}\) is similar in magnitude to the magnetic noise power at frequency \(\Delta_{0-}\). We conducted this comparison on NV1 with \(\Delta_{0-} = 2438 \ \text{MHz}\) and \(\Delta_{\pm} = 1017 \ \text{MHz}\). This data is marked with an asterisk in Table~S1 of the supplemental materials and shown in Fig.~S3 of the supplemental materials \footnotemark[2]. Using Eqs.~\ref{gamma_B} and \ref{Omega_B} along with numerically calculated eigenstates, we determine the ratio \(\Omega_{B, -}/\gamma_{B}\) to be \(\sim 25\) if we assume that the magnetic noise power is the same at \(\Delta_{\pm}\) and \(\Delta_{0-}\). We measure \(\Omega = 0.7(2) \ \text{kHz}\) on the \(\ket{H;0} \leftrightarrow \ket{H;-1}\) transition and \(\gamma = 0.41(10) \ \text{kHz}\). The ratio between these two rates is \(\sim 2\), or more than an order of magnitude below what would be expected if magnetic noise were the dominant source of relaxation. This indicates that magnetic noise combined with eigenstate mixing is not a significant source of \(\ket{H;-1} \leftrightarrow \ket{H;+1}\) relaxation. 

To further test whether or not the fast relaxation we observe is a result of a nonzero \(\gamma_{B}\) due to eigenstate mixing, we measured \(\gamma\) on NV1 at a splitting of \(\Delta_{\pm}=350 \ \text{MHz}\) after changing the composition of the eigenstates by rotating the applied magnetic field away from the optimally aligned angle at which the rest of the data for NV1 was recorded. This data is marked with a dagger in Table~S1 of the supplemental materials \footnotemark[2]. Using Eq. \ref{gamma_B} and numerically calculated eigenstates, we can quantitatively describe the ratio of \(\gamma_{B}\) in the rotated case to \(\gamma_{B}\) in the aligned case at the same splitting. We calculate this ratio to be \(\sim 3.5\). The measured value of \(\gamma\) in the rotated case is \(1.6(2) \ \text{kHz}\), whereas the value of \(\gamma\) extrapolated from the empirical fit shown in Fig.~\ref{fig:three} in the aligned case is \(1.02 \ \text{kHz}\). The ratio between these two rates is \(\sim 1.5\), which is comparable to other observed deviations from the fit in Fig.~\ref{fig:three}. This measured ratio is less than half the factor of \(\sim 3.5\) that we would expect if the relaxation were dominated by magnetic noise, providing additional evidence that magnetic noise is not the dominant source of qutrit relaxation.

\subsection{Attribution of fast relaxation rates \(\gamma\) to electric field noise}

While the numerical analysis described above discounts the possibility that magnetic noise is responsible for the fast qutrit relaxation rates, it does not provide an alternative explanation for our observations. The results of Kim, \emph{et al.}~\cite{Kim2015} and Jamonneau, \emph{et al.}~\cite{Jamonneau2016}, whose works analyzed the effect of electric field noise on the coherence times \(T_{2}\) and \(T_{2}^*\) respectively, suggest that electric field noise is to blame. Importantly, their methodologies are entirely distinct from the qutrit relaxation rate spectroscopy employed here and in Myers, \emph{et al.}~\cite{Myers2017}. Specifically, Kim, \emph{et al.}~compared the \(T_{2}\) times of shallow NVs before and after applying high dielectric liquid to the bulk diamond surface.
Jamonneau, \emph{et al.} quantified low frequency electric field and magnetic field noise by measuring \(T_{2}^*\) for energy eigenstates tuned to be protected against either magnetic field noise or electric field noise. Both groups found evidence of significant low frequency electric field noise, which supports an interpretation of qutrit relaxation rate measurements in nanodiamonds as spectroscopic probes of electric field noise, as has previously also been concluded for shallow NVs in bulk diamond by Myers, \emph{et al.}~\cite{Myers2017}.


Based on the above considerations, we attribute the fast $\gamma$ relaxation rates to resonant electric field noise at the frequency $\Delta_{\pm}$ incoherently driving transitions between $\ket{H;\pm1}$ through the $d_\perp$ terms in Eq.~\ref{hamiltonian_Pi} \footnotemark[2]. While our measurements cannot discriminate between strain and electric field noise, based on our observations of the noise changing in time as described below, and guided by other works where electric field noise emanating from surfaces was observed with NVs \cite{ Kim2015,Myers2017, Jamonneau2016, Sangtawesin2019} and in other materials systems \cite{devoret2004, Brownnutt2015, christensen2019,Kuhlmann2013}, we argue that the observed noise is primarily electric field noise emanating from the nanodiamond surfaces. Under this assumption, the measured relaxation rates are then directly proportional to the perpendicular electric field noise power spectrum: $S_{E_\perp}(\omega) = \gamma(\omega) / (d_{\perp}^2/h^2)$ \cite{Myers2017, Kim2015, Sangtawesin2019}. Using the functions of $\gamma$ determined in Fig.~\ref{fig:three}, we integrate $S_{E_\perp}(\omega)$ over the range of measured frequencies $\omega\approx20 - 1000$ MHz and find an order of magnitude estimate of $E^{RMS}_{\perp}=10^7$~V/m for all five NVs. For comparison, this value is roughly an order of magnitude larger than the d.c.~electric field 20 nm away from a single stationary electron.
The $1/f^{2}$ scaling we observe at lower splittings hints at fluctuating electric dipoles or the motion of charges between charge traps as possible sources of the noise \cite{Myers2017}. While our results, in combination with previous studies of the charge noise on diamond surfaces \cite{Myers2017, Kim2015, Safavi-Naini2011, Jamonneau2016}, provide some clues to its origin, the precise nature of the physical process remains an open question. Additionally, we note that \(\Omega_{avg}\) and \(\gamma_{\infty}\) are of the same order of magnitude for each of the NVs surveyed. This suggests that \(\Omega\) may also be limited by resonant electric field noise driving \(\ket{H;0} \leftrightarrow \ket{H;\pm1}\) transitions, and provides indirect evidence that \(d_{\perp}'\neq0\).

\subsection{Implications for coherence time \(T_{2}\)}

Critically, because the relaxation between $\ket{H;\pm1}$ is an incoherent process, it sets a hard limit on the achievable coherence time $T_2$ for an NV spin qubit formed from $\ket{H;0}$ and either $\ket{H;+1}$ or $\ket{H;-1}$ \cite{Myers2017} of
\begin{equation}\label{t2_max}
T_\text{2,max} = 2 \left(3\Omega + \gamma\right)^{-1}.
\end{equation} 
Using the measured values of $\gamma$ and $\Omega$, this value is plotted as a function of $\Delta_{\pm}$ in the insets of Fig.~\ref{fig:three}.  At low splitting, $T_\text{2,max}$ is severely limited by $\gamma$ and is $\sim2$ orders of magnitude shorter than the erroneous $T_\text{2,max}$ that would be calculated under the assumption that the NV spin is a qubit. 
These observations may help to explain the mixed results of prior attempts to use dynamical decoupling to extend the coherence times of NVs in nanodiamonds \cite{Naydenov2011, Knowles2014, Liu2014}. They also indicate that, whenever possible, measurements with NVs in nanodiamonds should be performed at axial magnetic fields exceeding $\sim60$ G. In addition, the observed $1/f^{2}$ noise scaling implies that the electric field noise power spectral density is even larger at lower frequencies, and as these slowly fluctuating fields will shift the energies of the $\ket{H;0} \leftrightarrow \ket{H;\pm1}$ transitions through the $d_{\parallel}\Pi_{z}S_{z}^{2}$ term in Eq.~\ref{hamiltonian_Pi}, low frequency charge noise must be considered as a significant source of dephasing in nanodiamonds. This is consistent with the measurements by Jamonneau \emph{et al.} of $T_2^*$ as a function of axial magnetic field performed on a single NV in a nanodiamond \cite{Jamonneau2016}. 

\begin{figure}[t]
\includegraphics{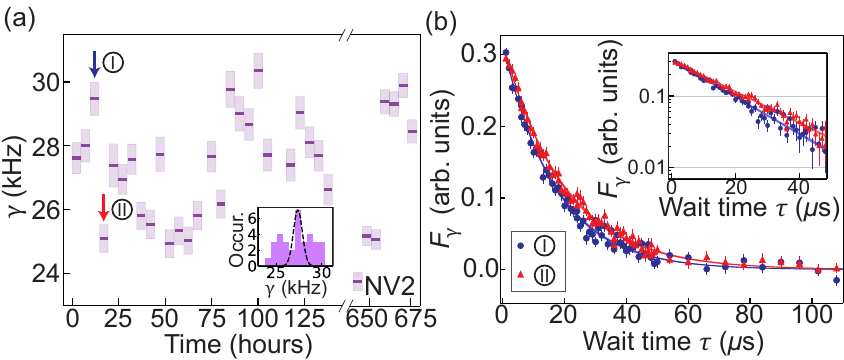}
\caption{\label{fig:four} Temporal fluctuations in the relaxation rate $\gamma$. (a) Rate \(\gamma\)  of NV2 at a single splitting $\Delta_{\pm} \approx 29$~MHz (referenced by a black arrow in Fig.~\ref{fig:three}(b)) measured for 140 consecutive hours, followed by a three week long gap, then measured again for 35 consecutive hours. Error bars are one standard error. Inset shows a histogram of the measured rates overlaid with a Gaussian curve with standard deviation equal to the average standard error of the measurements. (b) Measured relaxation curves $F_\gamma(\tau)$ (Eq.~\ref{gamma_subtracted_eq}) for the back-to-back measurements labeled in (a), illustrating a change in $\gamma$ that was well above our signal-to-noise ratio. The inset shows the first 50 $\mu$s of the same data plotted on a semi-log scale. Error bars are one standard error.}
\end{figure}


\subsection{Temporal fluctuations in \(\gamma\)}

As shown in Fig.~\ref{fig:three}, at several values for $\Delta_{\pm}$ we performed a measurement of \(\gamma\) multiple times and observed deviations well above our standard error. We attribute these deviations to variations in the local electric field noise power spectral density as a function of time.  Figure~\ref{fig:four} shows the value of $\gamma$ as a function of time by performing the same measurement repeatedly on NV2 at a constant splitting of $\sim$ 29 MHz. 
The changes in the rate $\gamma$ in Fig.~\ref{fig:four}(a) are well outside of the statistical uncertainty of the measurements, and indicate that $\gamma$ fluctuates in time over hours to days. Similar fluctuations were observed when the same measurements were performed with NV1 \footnotemark[2], and previously in shallow NVs in bulk diamond \cite{Myers2017,bluvstein2019}. While the origins of these temporal dynamics are presently obscure, they provide additional evidence that the observed noise is primarily from the nanodimaond surfaces, where adsorbates may be coming and going with time. 

\section{Conclusions}

In conclusion, we have presented observations of fast relaxation rates on qutrit transitions of NV electronic spins in nanodiamonds. We find that this relaxation rate depends on the splitting of the $\ket{H;\pm1}$ levels, and that the resulting limit on the coherence time $T_2$ improves with higher axial magnetic fields. Additionally, we observe this rate changing with time. Our results demonstrate that the qutrit relaxation rate $\gamma$ between the $\ket{H;\pm1}$ states is a critical figure of merit for the coherence of NVs in nanodiamonds, and suggest that electric field noise is a major source of both dephasing and relaxation. Beyond the scope of this work, future experiments could be performed to shed light on the origins of this noise and to develop methods for mitigating it, including studies of how $\gamma$ changes with surface functionalization, nanodiamond size, immersion in dielectric liquids, or temperature. Additionally, measurements of $\gamma$ could also be used as a local probe of the electric field noise near surfaces in quantum systems limited by charge noise, such as ion traps \cite{Brownnutt2015}, semiconductor quantum dots \cite{Kuhlmann2013}, and superconducting qubits \cite{devoret2004, christensen2019}.

\section*{Acknowledgments}

The authors thank Jeff Thompson, Nathalie de Leon, Jeronimo Maze, Ariel Norambuena, Ania Bleszynski Jayich, Mikhail Lukin, Dolev Bluvstein, and Norman Yao for enlightening discussions and helpful insights, and Wangping Ren and Sam Li for their contributions to the experimental apparatus. This work was supported by the U.S. Department of Energy, Office of Science, Basic Energy Sciences under Award \#DE-SC0020313. A.~G.~acknowledges support from the Department of Defense through the National Defense Science and Engineering Graduate Fellowship (NDSEG) program.

\bibliography{MainReferences2.bib}

\providecommand{\noopsort}[1]{}\providecommand{\singleletter}[1]{#1}%
\begin{thebibliography}{39}%
\makeatletter
\providecommand \@ifxundefined [1]{%
 \@ifx{#1\undefined}
}%
\providecommand \@ifnum [1]{%
 \ifnum #1\expandafter \@firstoftwo
 \else \expandafter \@secondoftwo
 \fi
}%
\providecommand \@ifx [1]{%
 \ifx #1\expandafter \@firstoftwo
 \else \expandafter \@secondoftwo
 \fi
}%
\providecommand \natexlab [1]{#1}%
\providecommand \enquote  [1]{``#1''}%
\providecommand \bibnamefont  [1]{#1}%
\providecommand \bibfnamefont [1]{#1}%
\providecommand \citenamefont [1]{#1}%
\providecommand \href@noop [0]{\@secondoftwo}%
\providecommand \href [0]{\begingroup \@sanitize@url \@href}%
\providecommand \@href[1]{\@@startlink{#1}\@@href}%
\providecommand \@@href[1]{\endgroup#1\@@endlink}%
\providecommand \@sanitize@url [0]{\catcode `\\12\catcode `\$12\catcode
  `\&12\catcode `\#12\catcode `\^12\catcode `\_12\catcode `\%12\relax}%
\providecommand \@@startlink[1]{}%
\providecommand \@@endlink[0]{}%
\providecommand \url  [0]{\begingroup\@sanitize@url \@url }%
\providecommand \@url [1]{\endgroup\@href {#1}{\urlprefix }}%
\providecommand \urlprefix  [0]{URL }%
\providecommand \Eprint [0]{\href }%
\providecommand \doibase [0]{https://doi.org/}%
\providecommand \selectlanguage [0]{\@gobble}%
\providecommand \bibinfo  [0]{\@secondoftwo}%
\providecommand \bibfield  [0]{\@secondoftwo}%
\providecommand \translation [1]{[#1]}%
\providecommand \BibitemOpen [0]{}%
\providecommand \bibitemStop [0]{}%
\providecommand \bibitemNoStop [0]{.\EOS\space}%
\providecommand \EOS [0]{\spacefactor3000\relax}%
\providecommand \BibitemShut  [1]{\csname bibitem#1\endcsname}%
\let\auto@bib@innerbib\@empty
\bibitem [{\citenamefont {Balasubramanian}\ \emph {et~al.}(2009)\citenamefont
  {Balasubramanian}, \citenamefont {Neumann}, \citenamefont {Twitchen},
  \citenamefont {Markham}, \citenamefont {Kolesov}, \citenamefont {Mizuochi},
  \citenamefont {Isoya}, \citenamefont {Achard}, \citenamefont {Beck},
  \citenamefont {Tissler} \emph {et~al.}}]{balasubramanian2009}%
  \BibitemOpen
  \bibfield  {author} {\bibinfo {author} {\bibfnamefont {G.}~\bibnamefont
  {Balasubramanian}}, \bibinfo {author} {\bibfnamefont {P.}~\bibnamefont
  {Neumann}}, \bibinfo {author} {\bibfnamefont {D.}~\bibnamefont {Twitchen}},
  \bibinfo {author} {\bibfnamefont {M.}~\bibnamefont {Markham}}, \bibinfo
  {author} {\bibfnamefont {R.}~\bibnamefont {Kolesov}}, \bibinfo {author}
  {\bibfnamefont {N.}~\bibnamefont {Mizuochi}}, \bibinfo {author}
  {\bibfnamefont {J.}~\bibnamefont {Isoya}}, \bibinfo {author} {\bibfnamefont
  {J.}~\bibnamefont {Achard}}, \bibinfo {author} {\bibfnamefont
  {J.}~\bibnamefont {Beck}}, \bibinfo {author} {\bibfnamefont {J.}~\bibnamefont
  {Tissler}}, \emph {et~al.},\ }\bibfield  {title} {\bibinfo {title} {Ultralong
  spin coherence time in isotopically engineered diamond},\ }\href@noop {}
  {\bibfield  {journal} {\bibinfo  {journal} {Nat. Mater.}\ }\textbf {\bibinfo
  {volume} {8}},\ \bibinfo {pages} {383} (\bibinfo {year} {2009})}\BibitemShut
  {NoStop}%
\bibitem [{\citenamefont {De~Lange}\ \emph {et~al.}(2010)\citenamefont
  {De~Lange}, \citenamefont {Wang}, \citenamefont {Riste}, \citenamefont
  {Dobrovitski},\ and\ \citenamefont {Hanson}}]{deLange2010universal}%
  \BibitemOpen
  \bibfield  {author} {\bibinfo {author} {\bibfnamefont {G.}~\bibnamefont
  {De~Lange}}, \bibinfo {author} {\bibfnamefont {Z.}~\bibnamefont {Wang}},
  \bibinfo {author} {\bibfnamefont {D.}~\bibnamefont {Riste}}, \bibinfo
  {author} {\bibfnamefont {V.}~\bibnamefont {Dobrovitski}},\ and\ \bibinfo
  {author} {\bibfnamefont {R.}~\bibnamefont {Hanson}},\ }\bibfield  {title}
  {\bibinfo {title} {Universal dynamical decoupling of a single solid-state
  spin from a spin bath},\ }\href@noop {} {\bibfield  {journal} {\bibinfo
  {journal} {Science}\ }\textbf {\bibinfo {volume} {330}},\ \bibinfo {pages}
  {60} (\bibinfo {year} {2010})}\BibitemShut {NoStop}%
\bibitem [{\citenamefont {Naydenov}\ \emph {et~al.}(2011)\citenamefont
  {Naydenov}, \citenamefont {Dolde}, \citenamefont {Hall}, \citenamefont
  {Shin}, \citenamefont {Fedder}, \citenamefont {Hollenberg}, \citenamefont
  {Jelezko},\ and\ \citenamefont {Wrachtrup}}]{Naydenov2011}%
  \BibitemOpen
  \bibfield  {author} {\bibinfo {author} {\bibfnamefont {B.}~\bibnamefont
  {Naydenov}}, \bibinfo {author} {\bibfnamefont {F.}~\bibnamefont {Dolde}},
  \bibinfo {author} {\bibfnamefont {L.~T.}\ \bibnamefont {Hall}}, \bibinfo
  {author} {\bibfnamefont {C.}~\bibnamefont {Shin}}, \bibinfo {author}
  {\bibfnamefont {H.}~\bibnamefont {Fedder}}, \bibinfo {author} {\bibfnamefont
  {L.~C.~L.}\ \bibnamefont {Hollenberg}}, \bibinfo {author} {\bibfnamefont
  {F.}~\bibnamefont {Jelezko}},\ and\ \bibinfo {author} {\bibfnamefont
  {J.}~\bibnamefont {Wrachtrup}},\ }\bibfield  {title} {\bibinfo {title}
  {Dynamical decoupling of a single-electron spin at room temperature},\
  }\href@noop {} {\bibfield  {journal} {\bibinfo  {journal} {Phys. Rev. B}\
  }\textbf {\bibinfo {volume} {83}},\ \bibinfo {pages} {081201} (\bibinfo
  {year} {2011})}\BibitemShut {NoStop}%
\bibitem [{\citenamefont {Bar-Gill}\ \emph {et~al.}(2013)\citenamefont
  {Bar-Gill}, \citenamefont {Pham}, \citenamefont {Jarmola}, \citenamefont
  {Budker},\ and\ \citenamefont {Walsworth}}]{bar_gill2013}%
  \BibitemOpen
  \bibfield  {author} {\bibinfo {author} {\bibfnamefont {N.}~\bibnamefont
  {Bar-Gill}}, \bibinfo {author} {\bibfnamefont {L.}~\bibnamefont {Pham}},
  \bibinfo {author} {\bibfnamefont {A.}~\bibnamefont {Jarmola}}, \bibinfo
  {author} {\bibfnamefont {D.}~\bibnamefont {Budker}},\ and\ \bibinfo {author}
  {\bibfnamefont {R.}~\bibnamefont {Walsworth}},\ }\bibfield  {title} {\bibinfo
  {title} {Solid-state electronic spin coherence time approaching one second},\
  }\href@noop {} {\bibfield  {journal} {\bibinfo  {journal} {Nat. Commun.}\
  }\textbf {\bibinfo {volume} {4}},\ \bibinfo {pages} {1743} (\bibinfo {year}
  {2013})}\BibitemShut {NoStop}%
\bibitem [{\citenamefont {Rondin}\ \emph {et~al.}(2014)\citenamefont {Rondin},
  \citenamefont {Tetienne}, \citenamefont {Hingant}, \citenamefont {Roch},
  \citenamefont {Maletinsky}, \citenamefont {Jacques} \emph
  {et~al.}}]{Rondin2014}%
  \BibitemOpen
  \bibfield  {author} {\bibinfo {author} {\bibfnamefont {L.}~\bibnamefont
  {Rondin}}, \bibinfo {author} {\bibfnamefont {J.-P.}\ \bibnamefont
  {Tetienne}}, \bibinfo {author} {\bibfnamefont {T.}~\bibnamefont {Hingant}},
  \bibinfo {author} {\bibfnamefont {J.-F.}\ \bibnamefont {Roch}}, \bibinfo
  {author} {\bibfnamefont {P.}~\bibnamefont {Maletinsky}}, \bibinfo {author}
  {\bibfnamefont {V.}~\bibnamefont {Jacques}}, \emph {et~al.},\ }\bibfield
  {title} {\bibinfo {title} {Magnetometry with nitrogen-vacancy defects in
  diamond},\ }\href@noop {} {\bibfield  {journal} {\bibinfo  {journal} {Rep.
  Prog. Phys.}\ }\textbf {\bibinfo {volume} {77}},\ \bibinfo {pages} {056503}
  (\bibinfo {year} {2014})}\BibitemShut {NoStop}%
\bibitem [{\citenamefont {Dolde}\ \emph {et~al.}(2011)\citenamefont {Dolde},
  \citenamefont {Fedder}, \citenamefont {Doherty}, \citenamefont {N{\"o}bauer},
  \citenamefont {Rempp}, \citenamefont {Balasubramanian}, \citenamefont {Wolf},
  \citenamefont {Reinhard}, \citenamefont {Hollenberg}, \citenamefont
  {Jelezko}, \citenamefont {Wrachtrup} \emph {et~al.}}]{Dolde2011}%
  \BibitemOpen
  \bibfield  {author} {\bibinfo {author} {\bibfnamefont {F.}~\bibnamefont
  {Dolde}}, \bibinfo {author} {\bibfnamefont {H.}~\bibnamefont {Fedder}},
  \bibinfo {author} {\bibfnamefont {M.~W.}\ \bibnamefont {Doherty}}, \bibinfo
  {author} {\bibfnamefont {T.}~\bibnamefont {N{\"o}bauer}}, \bibinfo {author}
  {\bibfnamefont {F.}~\bibnamefont {Rempp}}, \bibinfo {author} {\bibfnamefont
  {G.}~\bibnamefont {Balasubramanian}}, \bibinfo {author} {\bibfnamefont
  {T.}~\bibnamefont {Wolf}}, \bibinfo {author} {\bibfnamefont {F.}~\bibnamefont
  {Reinhard}}, \bibinfo {author} {\bibfnamefont {L.~C.~L.}\ \bibnamefont
  {Hollenberg}}, \bibinfo {author} {\bibfnamefont {F.}~\bibnamefont {Jelezko}},
  \bibinfo {author} {\bibfnamefont {J.}~\bibnamefont {Wrachtrup}}, \emph
  {et~al.},\ }\bibfield  {title} {\bibinfo {title} {Electric-field sensing
  using single diamond spins},\ }\href@noop {} {\bibfield  {journal} {\bibinfo
  {journal} {Nat. Phys.}\ }\textbf {\bibinfo {volume} {7}},\ \bibinfo {pages}
  {459} (\bibinfo {year} {2011})}\BibitemShut {NoStop}%
\bibitem [{\citenamefont {Mittiga}\ \emph {et~al.}(2018)\citenamefont
  {Mittiga}, \citenamefont {Hsieh}, \citenamefont {Zu}, \citenamefont {Kobrin},
  \citenamefont {Machado}, \citenamefont {Bhattacharyya}, \citenamefont {Rui},
  \citenamefont {Jarmola}, \citenamefont {Choi}, \citenamefont {Budker},
  \citenamefont {Yao} \emph {et~al.}}]{Mittiga2018}%
  \BibitemOpen
  \bibfield  {author} {\bibinfo {author} {\bibfnamefont {T.}~\bibnamefont
  {Mittiga}}, \bibinfo {author} {\bibfnamefont {S.}~\bibnamefont {Hsieh}},
  \bibinfo {author} {\bibfnamefont {C.}~\bibnamefont {Zu}}, \bibinfo {author}
  {\bibfnamefont {B.}~\bibnamefont {Kobrin}}, \bibinfo {author} {\bibfnamefont
  {F.}~\bibnamefont {Machado}}, \bibinfo {author} {\bibfnamefont
  {P.}~\bibnamefont {Bhattacharyya}}, \bibinfo {author} {\bibfnamefont {N.~Z.}\
  \bibnamefont {Rui}}, \bibinfo {author} {\bibfnamefont {A.}~\bibnamefont
  {Jarmola}}, \bibinfo {author} {\bibfnamefont {S.}~\bibnamefont {Choi}},
  \bibinfo {author} {\bibfnamefont {D.}~\bibnamefont {Budker}}, \bibinfo
  {author} {\bibfnamefont {N.~Y.}\ \bibnamefont {Yao}}, \emph {et~al.},\
  }\bibfield  {title} {\bibinfo {title} {Imaging the local charge environment
  of nitrogen-vacancy centers in diamond},\ }\href@noop {} {\bibfield
  {journal} {\bibinfo  {journal} {Phys. Rev. Lett.}\ }\textbf {\bibinfo
  {volume} {121}},\ \bibinfo {pages} {6} (\bibinfo {year} {2018})}\BibitemShut
  {NoStop}%
\bibitem [{\citenamefont {Ovartchaiyapong}\ \emph {et~al.}(2014)\citenamefont
  {Ovartchaiyapong}, \citenamefont {Lee}, \citenamefont {Myers},\ and\
  \citenamefont {Jayich}}]{Ovartchaiyapong2014}%
  \BibitemOpen
  \bibfield  {author} {\bibinfo {author} {\bibfnamefont {P.}~\bibnamefont
  {Ovartchaiyapong}}, \bibinfo {author} {\bibfnamefont {K.~W.}\ \bibnamefont
  {Lee}}, \bibinfo {author} {\bibfnamefont {B.~A.}\ \bibnamefont {Myers}},\
  and\ \bibinfo {author} {\bibfnamefont {A.~C.~B.}\ \bibnamefont {Jayich}},\
  }\bibfield  {title} {\bibinfo {title} {Dynamic strain-mediated coupling of a
  single diamond spin to a mechanical resonator},\ }\href@noop {} {\bibfield
  {journal} {\bibinfo  {journal} {Nat. Commun.}\ }\textbf {\bibinfo {volume}
  {5}},\ \bibinfo {pages} {4429} (\bibinfo {year} {2014})}\BibitemShut
  {NoStop}%
\bibitem [{\citenamefont {Teissier}\ \emph {et~al.}(2014)\citenamefont
  {Teissier}, \citenamefont {Barfuss}, \citenamefont {Appel}, \citenamefont
  {Neu},\ and\ \citenamefont {Maletinsky}}]{Teissier2014}%
  \BibitemOpen
  \bibfield  {author} {\bibinfo {author} {\bibfnamefont {J.}~\bibnamefont
  {Teissier}}, \bibinfo {author} {\bibfnamefont {A.}~\bibnamefont {Barfuss}},
  \bibinfo {author} {\bibfnamefont {P.}~\bibnamefont {Appel}}, \bibinfo
  {author} {\bibfnamefont {E.}~\bibnamefont {Neu}},\ and\ \bibinfo {author}
  {\bibfnamefont {P.}~\bibnamefont {Maletinsky}},\ }\bibfield  {title}
  {\bibinfo {title} {Strain coupling of a nitrogen-vacancy center spin to a
  diamond mechanical oscillator},\ }\href
  {https://doi.org/10.1103/PhysRevLett.113.020503} {\bibfield  {journal}
  {\bibinfo  {journal} {Phys. Rev. Lett.}\ }\textbf {\bibinfo {volume} {113}},\
  \bibinfo {pages} {020503} (\bibinfo {year} {2014})}\BibitemShut {NoStop}%
\bibitem [{\citenamefont {Kucsko}\ \emph {et~al.}(2013)\citenamefont {Kucsko},
  \citenamefont {Maurer}, \citenamefont {Yao}, \citenamefont {Kubo},
  \citenamefont {Noh}, \citenamefont {Lo}, \citenamefont {Park},\ and\
  \citenamefont {Lukin}}]{Kucsko2013}%
  \BibitemOpen
  \bibfield  {author} {\bibinfo {author} {\bibfnamefont {G.}~\bibnamefont
  {Kucsko}}, \bibinfo {author} {\bibfnamefont {P.~C.}\ \bibnamefont {Maurer}},
  \bibinfo {author} {\bibfnamefont {N.~Y.}\ \bibnamefont {Yao}}, \bibinfo
  {author} {\bibfnamefont {M.}~\bibnamefont {Kubo}}, \bibinfo {author}
  {\bibfnamefont {H.~J.}\ \bibnamefont {Noh}}, \bibinfo {author} {\bibfnamefont
  {P.~K.}\ \bibnamefont {Lo}}, \bibinfo {author} {\bibfnamefont
  {H.}~\bibnamefont {Park}},\ and\ \bibinfo {author} {\bibfnamefont {M.~D.}\
  \bibnamefont {Lukin}},\ }\bibfield  {title} {\bibinfo {title}
  {Nanometre-scale thermometry in a living cell},\ }\href@noop {} {\bibfield
  {journal} {\bibinfo  {journal} {Nature (London)}\ }\textbf {\bibinfo {volume}
  {500}},\ \bibinfo {pages} {54} (\bibinfo {year} {2013})}\BibitemShut
  {NoStop}%
\bibitem [{\citenamefont {Toyli}\ \emph {et~al.}(2013)\citenamefont {Toyli},
  \citenamefont {de~las Casas}, \citenamefont {Christle}, \citenamefont
  {Dobrovitski},\ and\ \citenamefont {Awschalom}}]{Toyli2013}%
  \BibitemOpen
  \bibfield  {author} {\bibinfo {author} {\bibfnamefont {D.~M.}\ \bibnamefont
  {Toyli}}, \bibinfo {author} {\bibfnamefont {C.~F.}\ \bibnamefont {de~las
  Casas}}, \bibinfo {author} {\bibfnamefont {D.~J.}\ \bibnamefont {Christle}},
  \bibinfo {author} {\bibfnamefont {V.~V.}\ \bibnamefont {Dobrovitski}},\ and\
  \bibinfo {author} {\bibfnamefont {D.~D.}\ \bibnamefont {Awschalom}},\
  }\bibfield  {title} {\bibinfo {title} {Fluorescence thermometry enhanced by
  the quantum coherence of single spins in diamond},\ }\href@noop {} {\bibfield
   {journal} {\bibinfo  {journal} {P. Natl. Acad. Sci. USA}\ }\textbf {\bibinfo
  {volume} {110}},\ \bibinfo {pages} {8417} (\bibinfo {year}
  {2013})}\BibitemShut {NoStop}%
\bibitem [{\citenamefont {Kr{\"u}ger}\ \emph {et~al.}(2006)\citenamefont
  {Kr{\"u}ger}, \citenamefont {Liang}, \citenamefont {Jarre},\ and\
  \citenamefont {Stegk}}]{Kruger2006}%
  \BibitemOpen
  \bibfield  {author} {\bibinfo {author} {\bibfnamefont {A.}~\bibnamefont
  {Kr{\"u}ger}}, \bibinfo {author} {\bibfnamefont {Y.}~\bibnamefont {Liang}},
  \bibinfo {author} {\bibfnamefont {G.}~\bibnamefont {Jarre}},\ and\ \bibinfo
  {author} {\bibfnamefont {J.}~\bibnamefont {Stegk}},\ }\bibfield  {title}
  {\bibinfo {title} {Surface functionalisation of detonation diamond suitable
  for biological applications},\ }\href@noop {} {\bibfield  {journal} {\bibinfo
   {journal} {J. Mater. Chem.}\ }\textbf {\bibinfo {volume} {16}},\ \bibinfo
  {pages} {2322} (\bibinfo {year} {2006})}\BibitemShut {NoStop}%
\bibitem [{\citenamefont {Balasubramanian}\ \emph {et~al.}(2008)\citenamefont
  {Balasubramanian}, \citenamefont {Chan}, \citenamefont {Kolesov},
  \citenamefont {Al-Hmoud}, \citenamefont {Tisler}, \citenamefont {Shin},
  \citenamefont {Kim}, \citenamefont {Wojcik}, \citenamefont {Hemmer},
  \citenamefont {Krueger} \emph {et~al.}}]{Balasubramanian2008}%
  \BibitemOpen
  \bibfield  {author} {\bibinfo {author} {\bibfnamefont {G.}~\bibnamefont
  {Balasubramanian}}, \bibinfo {author} {\bibfnamefont {I.}~\bibnamefont
  {Chan}}, \bibinfo {author} {\bibfnamefont {R.}~\bibnamefont {Kolesov}},
  \bibinfo {author} {\bibfnamefont {M.}~\bibnamefont {Al-Hmoud}}, \bibinfo
  {author} {\bibfnamefont {J.}~\bibnamefont {Tisler}}, \bibinfo {author}
  {\bibfnamefont {C.}~\bibnamefont {Shin}}, \bibinfo {author} {\bibfnamefont
  {C.}~\bibnamefont {Kim}}, \bibinfo {author} {\bibfnamefont {A.}~\bibnamefont
  {Wojcik}}, \bibinfo {author} {\bibfnamefont {P.~R.}\ \bibnamefont {Hemmer}},
  \bibinfo {author} {\bibfnamefont {A.}~\bibnamefont {Krueger}}, \emph
  {et~al.},\ }\bibfield  {title} {\bibinfo {title} {Nanoscale imaging
  magnetometry with diamond spins under ambient conditions},\ }\href@noop {}
  {\bibfield  {journal} {\bibinfo  {journal} {Nature (London)}\ }\textbf
  {\bibinfo {volume} {455}},\ \bibinfo {pages} {648} (\bibinfo {year}
  {2008})}\BibitemShut {NoStop}%
\bibitem [{\citenamefont {Bogdanov}\ \emph {et~al.}(2019)\citenamefont
  {Bogdanov}, \citenamefont {Makarova}, \citenamefont {Lagutchev},
  \citenamefont {Shah}, \citenamefont {Chiang}, \citenamefont {Saha},
  \citenamefont {Baburin}, \citenamefont {Ryzhikov}, \citenamefont {Rodionov},
  \citenamefont {Kildishev}, \citenamefont {Boltasseva}, \citenamefont
  {Shalaev} \emph {et~al.}}]{Bogdanov2019}%
  \BibitemOpen
  \bibfield  {author} {\bibinfo {author} {\bibfnamefont {S.~I.}\ \bibnamefont
  {Bogdanov}}, \bibinfo {author} {\bibfnamefont {O.~A.}\ \bibnamefont
  {Makarova}}, \bibinfo {author} {\bibfnamefont {A.~S.}\ \bibnamefont
  {Lagutchev}}, \bibinfo {author} {\bibfnamefont {D.}~\bibnamefont {Shah}},
  \bibinfo {author} {\bibfnamefont {C.-C.}\ \bibnamefont {Chiang}}, \bibinfo
  {author} {\bibfnamefont {S.}~\bibnamefont {Saha}}, \bibinfo {author}
  {\bibfnamefont {A.~S.}\ \bibnamefont {Baburin}}, \bibinfo {author}
  {\bibfnamefont {I.~A.}\ \bibnamefont {Ryzhikov}}, \bibinfo {author}
  {\bibfnamefont {I.~A.}\ \bibnamefont {Rodionov}}, \bibinfo {author}
  {\bibfnamefont {A.~V.}\ \bibnamefont {Kildishev}}, \bibinfo {author}
  {\bibfnamefont {A.}~\bibnamefont {Boltasseva}}, \bibinfo {author}
  {\bibfnamefont {V.~M.}\ \bibnamefont {Shalaev}}, \emph {et~al.},\ }\bibfield
  {title} {\bibinfo {title} {Deterministic integration of single
  nitrogen-vacancy centers into nanopatch antennas},\ }\href@noop {} {\bibfield
   {journal} {\bibinfo  {journal} {arXiv:1902.05996}\ } (\bibinfo {year}
  {2019})}\BibitemShut {NoStop}%
\bibitem [{\citenamefont {Schietinger}\ \emph {et~al.}(2009)\citenamefont
  {Schietinger}, \citenamefont {Barth}, \citenamefont {Aichele},\ and\
  \citenamefont {Benson}}]{Schietinger2009}%
  \BibitemOpen
  \bibfield  {author} {\bibinfo {author} {\bibfnamefont {S.}~\bibnamefont
  {Schietinger}}, \bibinfo {author} {\bibfnamefont {M.}~\bibnamefont {Barth}},
  \bibinfo {author} {\bibfnamefont {T.}~\bibnamefont {Aichele}},\ and\ \bibinfo
  {author} {\bibfnamefont {O.}~\bibnamefont {Benson}},\ }\bibfield  {title}
  {\bibinfo {title} {Plasmon-enhanced single photon emission from a
  nanoassembled metal-diamond hybrid structure at room temperature},\
  }\href@noop {} {\bibfield  {journal} {\bibinfo  {journal} {Nano Lett.}\
  }\textbf {\bibinfo {volume} {9}},\ \bibinfo {pages} {1694 } (\bibinfo {year}
  {2009})}\BibitemShut {NoStop}%
\bibitem [{\citenamefont {Neukirch}\ \emph {et~al.}(2013)\citenamefont
  {Neukirch}, \citenamefont {Gieseler}, \citenamefont {Quidant}, \citenamefont
  {Novotny},\ and\ \citenamefont {Vamivakas}}]{Neukirch2013}%
  \BibitemOpen
  \bibfield  {author} {\bibinfo {author} {\bibfnamefont {L.~P.}\ \bibnamefont
  {Neukirch}}, \bibinfo {author} {\bibfnamefont {J.}~\bibnamefont {Gieseler}},
  \bibinfo {author} {\bibfnamefont {R.}~\bibnamefont {Quidant}}, \bibinfo
  {author} {\bibfnamefont {L.}~\bibnamefont {Novotny}},\ and\ \bibinfo {author}
  {\bibfnamefont {A.~N.}\ \bibnamefont {Vamivakas}},\ }\bibfield  {title}
  {\bibinfo {title} {Observation of nitrogen vacancy photoluminescence from an
  optically levitated nanodiamond},\ }\href@noop {} {\bibfield  {journal}
  {\bibinfo  {journal} {Opt. Lett.}\ }\textbf {\bibinfo {volume} {38}},\
  \bibinfo {pages} {2976} (\bibinfo {year} {2013})}\BibitemShut {NoStop}%
\bibitem [{\citenamefont {Hoang}\ \emph {et~al.}(2016)\citenamefont {Hoang},
  \citenamefont {Ahn}, \citenamefont {Bang},\ and\ \citenamefont
  {Li}}]{Hoang2016}%
  \BibitemOpen
  \bibfield  {author} {\bibinfo {author} {\bibfnamefont {T.~M.}\ \bibnamefont
  {Hoang}}, \bibinfo {author} {\bibfnamefont {J.}~\bibnamefont {Ahn}}, \bibinfo
  {author} {\bibfnamefont {J.}~\bibnamefont {Bang}},\ and\ \bibinfo {author}
  {\bibfnamefont {T.}~\bibnamefont {Li}},\ }\bibfield  {title} {\bibinfo
  {title} {Electron spin control of optically levitated nanodiamonds in
  vacuum},\ }\href@noop {} {\bibfield  {journal} {\bibinfo  {journal} {Nat.
  Commun.}\ }\textbf {\bibinfo {volume} {7}} (\bibinfo {year}
  {2016})}\BibitemShut {NoStop}%
\bibitem [{\citenamefont {Song}\ \emph {et~al.}(2014)\citenamefont {Song},
  \citenamefont {Zhang}, \citenamefont {Feng}, \citenamefont {Wang},
  \citenamefont {Zhang}, \citenamefont {Lou}, \citenamefont {Zhu},\ and\
  \citenamefont {Wang}}]{Song2014}%
  \BibitemOpen
  \bibfield  {author} {\bibinfo {author} {\bibfnamefont {X.}~\bibnamefont
  {Song}}, \bibinfo {author} {\bibfnamefont {J.}~\bibnamefont {Zhang}},
  \bibinfo {author} {\bibfnamefont {F.}~\bibnamefont {Feng}}, \bibinfo {author}
  {\bibfnamefont {J.}~\bibnamefont {Wang}}, \bibinfo {author} {\bibfnamefont
  {W.}~\bibnamefont {Zhang}}, \bibinfo {author} {\bibfnamefont
  {L.}~\bibnamefont {Lou}}, \bibinfo {author} {\bibfnamefont {W.}~\bibnamefont
  {Zhu}},\ and\ \bibinfo {author} {\bibfnamefont {G.}~\bibnamefont {Wang}},\
  }\bibfield  {title} {\bibinfo {title} {A statistical correlation
  investigation for the role of surface spins to the spin relaxation of
  nitrogen vacancy centers},\ }\href@noop {} {\bibfield  {journal} {\bibinfo
  {journal} {AIP Adv.}\ }\textbf {\bibinfo {volume} {4}},\ \bibinfo {pages}
  {047103} (\bibinfo {year} {2014})}\BibitemShut {NoStop}%
\bibitem [{\citenamefont {Knowles}\ \emph {et~al.}(2014)\citenamefont
  {Knowles}, \citenamefont {Kara},\ and\ \citenamefont
  {Atat{\"u}re}}]{Knowles2014}%
  \BibitemOpen
  \bibfield  {author} {\bibinfo {author} {\bibfnamefont {H.~S.}\ \bibnamefont
  {Knowles}}, \bibinfo {author} {\bibfnamefont {D.~M.}\ \bibnamefont {Kara}},\
  and\ \bibinfo {author} {\bibfnamefont {M.}~\bibnamefont {Atat{\"u}re}},\
  }\bibfield  {title} {\bibinfo {title} {Observing bulk diamond spin coherence
  in high-purity nanodiamonds},\ }\href@noop {} {\bibfield  {journal} {\bibinfo
   {journal} {Nat. Mater.}\ }\textbf {\bibinfo {volume} {13}},\ \bibinfo
  {pages} {21} (\bibinfo {year} {2014})}\BibitemShut {NoStop}%
\bibitem [{\citenamefont {Tsukahara}\ \emph {et~al.}(2019)\citenamefont
  {Tsukahara}, \citenamefont {Fujiwara}, \citenamefont {Sera}, \citenamefont
  {Nishimura}, \citenamefont {Sugai}, \citenamefont {Jentgens}, \citenamefont
  {Teki}, \citenamefont {Hashimoto},\ and\ \citenamefont
  {Shikata}}]{tsukahara2019}%
  \BibitemOpen
  \bibfield  {author} {\bibinfo {author} {\bibfnamefont {R.}~\bibnamefont
  {Tsukahara}}, \bibinfo {author} {\bibfnamefont {M.}~\bibnamefont {Fujiwara}},
  \bibinfo {author} {\bibfnamefont {Y.}~\bibnamefont {Sera}}, \bibinfo {author}
  {\bibfnamefont {Y.}~\bibnamefont {Nishimura}}, \bibinfo {author}
  {\bibfnamefont {Y.}~\bibnamefont {Sugai}}, \bibinfo {author} {\bibfnamefont
  {C.}~\bibnamefont {Jentgens}}, \bibinfo {author} {\bibfnamefont
  {Y.}~\bibnamefont {Teki}}, \bibinfo {author} {\bibfnamefont {H.}~\bibnamefont
  {Hashimoto}},\ and\ \bibinfo {author} {\bibfnamefont {S.}~\bibnamefont
  {Shikata}},\ }\bibfield  {title} {\bibinfo {title} {Removing
  non-size-dependent electron spin decoherence of nanodiamond quantum sensors
  by aerobic oxidation},\ }\href@noop {} {\bibfield  {journal} {\bibinfo
  {journal} {ACS Appl. Nano Mater.}\ } (\bibinfo {year} {2019})}\BibitemShut
  {NoStop}%
\bibitem [{\citenamefont {Liu}\ \emph {et~al.}(2014)\citenamefont {Liu},
  \citenamefont {Liu}, \citenamefont {Chang},\ and\ \citenamefont
  {Pan}}]{Liu2014}%
  \BibitemOpen
  \bibfield  {author} {\bibinfo {author} {\bibfnamefont {D.-Q.}\ \bibnamefont
  {Liu}}, \bibinfo {author} {\bibfnamefont {G.-Q.}\ \bibnamefont {Liu}},
  \bibinfo {author} {\bibfnamefont {Y.-C.}\ \bibnamefont {Chang}},\ and\
  \bibinfo {author} {\bibfnamefont {X.-Y.}\ \bibnamefont {Pan}},\ }\bibfield
  {title} {\bibinfo {title} {Scaling of dynamical decoupling for a single
  electron spin in nanodiamonds at room temperature},\ }\href@noop {}
  {\bibfield  {journal} {\bibinfo  {journal} {Physica B}\ }\textbf {\bibinfo
  {volume} {432}},\ \bibinfo {pages} {84 } (\bibinfo {year}
  {2014})}\BibitemShut {NoStop}%
\bibitem [{\citenamefont {Brandenburg}\ \emph {et~al.}(2018)\citenamefont
  {Brandenburg}, \citenamefont {Nagumo}, \citenamefont {Saichi}, \citenamefont
  {Tahara}, \citenamefont {Iwasaki}, \citenamefont {Hatano}, \citenamefont
  {Jelezko}, \citenamefont {Igarashi},\ and\ \citenamefont
  {Yatsui}}]{Brandenburg2018}%
  \BibitemOpen
  \bibfield  {author} {\bibinfo {author} {\bibfnamefont {F.}~\bibnamefont
  {Brandenburg}}, \bibinfo {author} {\bibfnamefont {R.}~\bibnamefont {Nagumo}},
  \bibinfo {author} {\bibfnamefont {K.}~\bibnamefont {Saichi}}, \bibinfo
  {author} {\bibfnamefont {K.}~\bibnamefont {Tahara}}, \bibinfo {author}
  {\bibfnamefont {T.}~\bibnamefont {Iwasaki}}, \bibinfo {author} {\bibfnamefont
  {M.}~\bibnamefont {Hatano}}, \bibinfo {author} {\bibfnamefont
  {F.}~\bibnamefont {Jelezko}}, \bibinfo {author} {\bibfnamefont
  {R.}~\bibnamefont {Igarashi}},\ and\ \bibinfo {author} {\bibfnamefont
  {T.}~\bibnamefont {Yatsui}},\ }\bibfield  {title} {\bibinfo {title}
  {Improving the electron spin properties of nitrogen-vacancy centres in
  nanodiamonds by near-field etching},\ }\href@noop {} {\bibfield  {journal}
  {\bibinfo  {journal} {Sci. Rep.}\ }\textbf {\bibinfo {volume} {8}},\ \bibinfo
  {pages} {15847} (\bibinfo {year} {2018})}\BibitemShut {NoStop}%
\bibitem [{\citenamefont {Ryan}\ \emph {et~al.}(2018)\citenamefont {Ryan},
  \citenamefont {Stacey}, \citenamefont {O'Donnell}, \citenamefont {Ohshima},
  \citenamefont {Johnson}, \citenamefont {Hollenberg}, \citenamefont
  {Mulvaney},\ and\ \citenamefont {Simpson}}]{Ryan2018}%
  \BibitemOpen
  \bibfield  {author} {\bibinfo {author} {\bibfnamefont {R.}~\bibnamefont
  {Ryan}}, \bibinfo {author} {\bibfnamefont {A.}~\bibnamefont {Stacey}},
  \bibinfo {author} {\bibfnamefont {K.~M.}\ \bibnamefont {O'Donnell}}, \bibinfo
  {author} {\bibfnamefont {T.}~\bibnamefont {Ohshima}}, \bibinfo {author}
  {\bibfnamefont {B.~C.}\ \bibnamefont {Johnson}}, \bibinfo {author}
  {\bibfnamefont {L.~C.~L.}\ \bibnamefont {Hollenberg}}, \bibinfo {author}
  {\bibfnamefont {P.}~\bibnamefont {Mulvaney}},\ and\ \bibinfo {author}
  {\bibfnamefont {D.~A.}\ \bibnamefont {Simpson}},\ }\bibfield  {title}
  {\bibinfo {title} {Impact of surface functionalization on the quantum
  coherence of nitrogen-vacancy centers in nanodiamonds},\ }\href@noop {}
  {\bibfield  {journal} {\bibinfo  {journal} {ACS Appl. Mater. Inter.}\
  }\textbf {\bibinfo {volume} {10}},\ \bibinfo {pages} {13143 } (\bibinfo
  {year} {2018})}\BibitemShut {NoStop}%
\bibitem [{\citenamefont {Myers}\ \emph {et~al.}(2017)\citenamefont {Myers},
  \citenamefont {Ariyaratne},\ and\ \citenamefont {Jayich}}]{Myers2017}%
  \BibitemOpen
  \bibfield  {author} {\bibinfo {author} {\bibfnamefont {B.~A.}\ \bibnamefont
  {Myers}}, \bibinfo {author} {\bibfnamefont {A.}~\bibnamefont {Ariyaratne}},\
  and\ \bibinfo {author} {\bibfnamefont {A.~C.~B.}\ \bibnamefont {Jayich}},\
  }\bibfield  {title} {\bibinfo {title} {Double-quantum spin-relaxation limits
  to coherence of near-surface nitrogen-vacancy centers},\ }\href@noop {}
  {\bibfield  {journal} {\bibinfo  {journal} {Phys. Rev. Lett.}\ }\textbf
  {\bibinfo {volume} {118}},\ \bibinfo {pages} {7} (\bibinfo {year}
  {2017})}\BibitemShut {NoStop}%
\bibitem [{Note1()}]{Note1}%
  \BibitemOpen
  \bibinfo {note} {See the product details from Ad\'{a}mas Nano for more
  information about these nanodiamonds with item number
  NDNV40nmLw10ml}\BibitemShut {NoStop}%
\bibitem [{Note2()}]{Note2}%
  \BibitemOpen
  \bibinfo {note} {See supplemental material for experimental details and
  additional data.}\BibitemShut {Stop}%
\bibitem [{\citenamefont {Kolkowitz}\ \emph {et~al.}(2015)\citenamefont
  {Kolkowitz}, \citenamefont {Safira}, \citenamefont {High}, \citenamefont
  {Devlin}, \citenamefont {Choi}, \citenamefont {Unterreithmeier},
  \citenamefont {Patterson}, \citenamefont {Zibrov}, \citenamefont
  {Manucharyan}, \citenamefont {Park}, \citenamefont {Lukin} \emph
  {et~al.}}]{Kolkowitz2015}%
  \BibitemOpen
  \bibfield  {author} {\bibinfo {author} {\bibfnamefont {S.}~\bibnamefont
  {Kolkowitz}}, \bibinfo {author} {\bibfnamefont {A.}~\bibnamefont {Safira}},
  \bibinfo {author} {\bibfnamefont {A.~A.}\ \bibnamefont {High}}, \bibinfo
  {author} {\bibfnamefont {R.~C.}\ \bibnamefont {Devlin}}, \bibinfo {author}
  {\bibfnamefont {S.}~\bibnamefont {Choi}}, \bibinfo {author} {\bibfnamefont
  {Q.~P.}\ \bibnamefont {Unterreithmeier}}, \bibinfo {author} {\bibfnamefont
  {D.}~\bibnamefont {Patterson}}, \bibinfo {author} {\bibfnamefont {A.~S.}\
  \bibnamefont {Zibrov}}, \bibinfo {author} {\bibfnamefont {V.~E.}\
  \bibnamefont {Manucharyan}}, \bibinfo {author} {\bibfnamefont
  {H.}~\bibnamefont {Park}}, \bibinfo {author} {\bibfnamefont {M.~D.}\
  \bibnamefont {Lukin}}, \emph {et~al.},\ }\bibfield  {title} {\bibinfo {title}
  {Probing {J}ohnson noise and ballistic transport in normal metals with a
  single-spin qubit},\ }\href@noop {} {\bibfield  {journal} {\bibinfo
  {journal} {Science}\ }\textbf {\bibinfo {volume} {347}},\ \bibinfo {pages}
  {1129} (\bibinfo {year} {2015})}\BibitemShut {NoStop}%
\bibitem [{\citenamefont {Sangtawesin}\ \emph {et~al.}(2019)\citenamefont
  {Sangtawesin}, \citenamefont {Dwyer}, \citenamefont {Srinivasan},
  \citenamefont {Allred}, \citenamefont {Rodgers}, \citenamefont {De~Greve},
  \citenamefont {Stacey}, \citenamefont {Dontschuk}, \citenamefont {O'Donnell},
  \citenamefont {Hu}, \citenamefont {Evans}, \citenamefont {Jaye},
  \citenamefont {Fischer}, \citenamefont {Markham}, \citenamefont {Twitchen},
  \citenamefont {Park}, \citenamefont {Lukin}, \citenamefont {de~Leon} \emph
  {et~al.}}]{Sangtawesin2019}%
  \BibitemOpen
  \bibfield  {author} {\bibinfo {author} {\bibfnamefont {S.}~\bibnamefont
  {Sangtawesin}}, \bibinfo {author} {\bibfnamefont {B.~L.}\ \bibnamefont
  {Dwyer}}, \bibinfo {author} {\bibfnamefont {S.}~\bibnamefont {Srinivasan}},
  \bibinfo {author} {\bibfnamefont {J.~J.}\ \bibnamefont {Allred}}, \bibinfo
  {author} {\bibfnamefont {L.~V.~H.}\ \bibnamefont {Rodgers}}, \bibinfo
  {author} {\bibfnamefont {K.}~\bibnamefont {De~Greve}}, \bibinfo {author}
  {\bibfnamefont {A.}~\bibnamefont {Stacey}}, \bibinfo {author} {\bibfnamefont
  {N.}~\bibnamefont {Dontschuk}}, \bibinfo {author} {\bibfnamefont {K.~M.}\
  \bibnamefont {O'Donnell}}, \bibinfo {author} {\bibfnamefont {D.}~\bibnamefont
  {Hu}}, \bibinfo {author} {\bibfnamefont {D.~A.}\ \bibnamefont {Evans}},
  \bibinfo {author} {\bibfnamefont {C.}~\bibnamefont {Jaye}}, \bibinfo {author}
  {\bibfnamefont {D.~A.}\ \bibnamefont {Fischer}}, \bibinfo {author}
  {\bibfnamefont {M.~L.}\ \bibnamefont {Markham}}, \bibinfo {author}
  {\bibfnamefont {D.~J.}\ \bibnamefont {Twitchen}}, \bibinfo {author}
  {\bibfnamefont {H.}~\bibnamefont {Park}}, \bibinfo {author} {\bibfnamefont
  {M.~D.}\ \bibnamefont {Lukin}}, \bibinfo {author} {\bibfnamefont {N.~P.}\
  \bibnamefont {de~Leon}}, \emph {et~al.},\ }\bibfield  {title} {\bibinfo
  {title} {Origins of diamond surface noise probed by correlating single-spin
  measurements with surface spectroscopy},\ }\href
  {https://doi.org/10.1103/PhysRevX.9.031052} {\bibfield  {journal} {\bibinfo
  {journal} {Phys. Rev. X}\ }\textbf {\bibinfo {volume} {9}},\ \bibinfo {pages}
  {031052} (\bibinfo {year} {2019})}\BibitemShut {NoStop}%
\bibitem [{\citenamefont {Ariyaratne}\ \emph {et~al.}(2018)\citenamefont
  {Ariyaratne}, \citenamefont {Bluvstein}, \citenamefont {Myers},\ and\
  \citenamefont {Jayich}}]{Ariyaratne2018}%
  \BibitemOpen
  \bibfield  {author} {\bibinfo {author} {\bibfnamefont {A.}~\bibnamefont
  {Ariyaratne}}, \bibinfo {author} {\bibfnamefont {D.}~\bibnamefont
  {Bluvstein}}, \bibinfo {author} {\bibfnamefont {B.~A.}\ \bibnamefont
  {Myers}},\ and\ \bibinfo {author} {\bibfnamefont {A.~C.~B.}\ \bibnamefont
  {Jayich}},\ }\bibfield  {title} {\bibinfo {title} {Nanoscale electrical
  conductivity imaging using a nitrogen-vacancy center in diamond},\
  }\href@noop {} {\bibfield  {journal} {\bibinfo  {journal} {Nat. Commun.}\
  }\textbf {\bibinfo {volume} {9}},\ \bibinfo {pages} {2406} (\bibinfo {year}
  {2018})}\BibitemShut {NoStop}%
\bibitem [{\citenamefont {Doherty}\ \emph {et~al.}(2012)\citenamefont
  {Doherty}, \citenamefont {Dolde}, \citenamefont {Fedder}, \citenamefont
  {Jelezko}, \citenamefont {Wrachtrup}, \citenamefont {Manson},\ and\
  \citenamefont {Hollenberg}}]{Doherty2012}%
  \BibitemOpen
  \bibfield  {author} {\bibinfo {author} {\bibfnamefont {M.~W.}\ \bibnamefont
  {Doherty}}, \bibinfo {author} {\bibfnamefont {F.}~\bibnamefont {Dolde}},
  \bibinfo {author} {\bibfnamefont {H.}~\bibnamefont {Fedder}}, \bibinfo
  {author} {\bibfnamefont {F.}~\bibnamefont {Jelezko}}, \bibinfo {author}
  {\bibfnamefont {J.}~\bibnamefont {Wrachtrup}}, \bibinfo {author}
  {\bibfnamefont {N.~B.}\ \bibnamefont {Manson}},\ and\ \bibinfo {author}
  {\bibfnamefont {L.~C.~L.}\ \bibnamefont {Hollenberg}},\ }\bibfield  {title}
  {\bibinfo {title} {Theory of the ground-state spin of the
  {NV}${}^{\ensuremath{-}}$ center in diamond},\ }\href@noop {} {\bibfield
  {journal} {\bibinfo  {journal} {Phys. Rev. B}\ }\textbf {\bibinfo {volume}
  {85}},\ \bibinfo {pages} {205203} (\bibinfo {year} {2012})}\BibitemShut
  {NoStop}%
\bibitem [{\citenamefont {Oort}\ and\ \citenamefont
  {Glasbeek}(1990)}]{vanoort1990}%
  \BibitemOpen
  \bibfield  {author} {\bibinfo {author} {\bibfnamefont {E.~V.}\ \bibnamefont
  {Oort}}\ and\ \bibinfo {author} {\bibfnamefont {M.}~\bibnamefont
  {Glasbeek}},\ }\bibfield  {title} {\bibinfo {title} {Electric-field-induced
  modulation of spin echoes of {N-V} centers in diamond},\ }\href@noop {}
  {\bibfield  {journal} {\bibinfo  {journal} {Chem. Phys. Lett.}\ }\textbf
  {\bibinfo {volume} {168}},\ \bibinfo {pages} {529 } (\bibinfo {year}
  {1990})}\BibitemShut {NoStop}%
\bibitem [{\citenamefont {Kim}\ \emph {et~al.}(2015)\citenamefont {Kim},
  \citenamefont {Mamin}, \citenamefont {Sherwood}, \citenamefont {Ohno},
  \citenamefont {Awschalom},\ and\ \citenamefont {Rugar}}]{Kim2015}%
  \BibitemOpen
  \bibfield  {author} {\bibinfo {author} {\bibfnamefont {M.}~\bibnamefont
  {Kim}}, \bibinfo {author} {\bibfnamefont {H.~J.}\ \bibnamefont {Mamin}},
  \bibinfo {author} {\bibfnamefont {M.~H.}\ \bibnamefont {Sherwood}}, \bibinfo
  {author} {\bibfnamefont {K.}~\bibnamefont {Ohno}}, \bibinfo {author}
  {\bibfnamefont {D.~D.}\ \bibnamefont {Awschalom}},\ and\ \bibinfo {author}
  {\bibfnamefont {D.}~\bibnamefont {Rugar}},\ }\bibfield  {title} {\bibinfo
  {title} {Decoherence of near-surface nitrogen-vacancy centers due to electric
  field noise},\ }\href@noop {} {\bibfield  {journal} {\bibinfo  {journal}
  {Phys. Rev. Lett.}\ }\textbf {\bibinfo {volume} {115}},\ \bibinfo {pages}
  {087602} (\bibinfo {year} {2015})}\BibitemShut {NoStop}%
\bibitem [{\citenamefont {Jamonneau}\ \emph {et~al.}(2016)\citenamefont
  {Jamonneau}, \citenamefont {Lesik}, \citenamefont {Tetienne}, \citenamefont
  {Alvizu}, \citenamefont {Mayer}, \citenamefont {Dr\'eau}, \citenamefont
  {Kosen}, \citenamefont {Roch}, \citenamefont {Pezzagna}, \citenamefont
  {Meijer}, \citenamefont {Teraji}, \citenamefont {Kubo}, \citenamefont
  {Bertet}, \citenamefont {Maze}, \citenamefont {Jacques} \emph
  {et~al.}}]{Jamonneau2016}%
  \BibitemOpen
  \bibfield  {author} {\bibinfo {author} {\bibfnamefont {P.}~\bibnamefont
  {Jamonneau}}, \bibinfo {author} {\bibfnamefont {M.}~\bibnamefont {Lesik}},
  \bibinfo {author} {\bibfnamefont {J.~P.}\ \bibnamefont {Tetienne}}, \bibinfo
  {author} {\bibfnamefont {I.}~\bibnamefont {Alvizu}}, \bibinfo {author}
  {\bibfnamefont {L.}~\bibnamefont {Mayer}}, \bibinfo {author} {\bibfnamefont
  {A.}~\bibnamefont {Dr\'eau}}, \bibinfo {author} {\bibfnamefont
  {S.}~\bibnamefont {Kosen}}, \bibinfo {author} {\bibfnamefont {J.-F.}\
  \bibnamefont {Roch}}, \bibinfo {author} {\bibfnamefont {S.}~\bibnamefont
  {Pezzagna}}, \bibinfo {author} {\bibfnamefont {J.}~\bibnamefont {Meijer}},
  \bibinfo {author} {\bibfnamefont {T.}~\bibnamefont {Teraji}}, \bibinfo
  {author} {\bibfnamefont {Y.}~\bibnamefont {Kubo}}, \bibinfo {author}
  {\bibfnamefont {P.}~\bibnamefont {Bertet}}, \bibinfo {author} {\bibfnamefont
  {J.~R.}\ \bibnamefont {Maze}}, \bibinfo {author} {\bibfnamefont
  {V.}~\bibnamefont {Jacques}}, \emph {et~al.},\ }\href@noop {} {\bibfield
  {journal} {\bibinfo  {journal} {Phys. Rev. B}\ }\textbf {\bibinfo {volume}
  {93}},\ \bibinfo {pages} {024305} (\bibinfo {year} {2016})}\BibitemShut
  {NoStop}%
\bibitem [{\citenamefont {Devoret}\ \emph {et~al.}(2004)\citenamefont
  {Devoret}, \citenamefont {Wallraff},\ and\ \citenamefont
  {Martinis}}]{devoret2004}%
  \BibitemOpen
  \bibfield  {author} {\bibinfo {author} {\bibfnamefont {M.~H.}\ \bibnamefont
  {Devoret}}, \bibinfo {author} {\bibfnamefont {A.}~\bibnamefont {Wallraff}},\
  and\ \bibinfo {author} {\bibfnamefont {J.~M.}\ \bibnamefont {Martinis}},\
  }\bibfield  {title} {\bibinfo {title} {Superconducting qubits: A short
  review},\ }\href@noop {} {\bibfield  {journal} {\bibinfo  {journal}
  {arXiv:0411174}\ } (\bibinfo {year} {2004})}\BibitemShut {NoStop}%
\bibitem [{\citenamefont {Brownnutt}\ \emph {et~al.}(2015)\citenamefont
  {Brownnutt}, \citenamefont {Kumph}, \citenamefont {Rabl},\ and\ \citenamefont
  {Blatt}}]{Brownnutt2015}%
  \BibitemOpen
  \bibfield  {author} {\bibinfo {author} {\bibfnamefont {M.}~\bibnamefont
  {Brownnutt}}, \bibinfo {author} {\bibfnamefont {M.}~\bibnamefont {Kumph}},
  \bibinfo {author} {\bibfnamefont {P.}~\bibnamefont {Rabl}},\ and\ \bibinfo
  {author} {\bibfnamefont {R.}~\bibnamefont {Blatt}},\ }\href@noop {}
  {\bibfield  {journal} {\bibinfo  {journal} {Rev. Mod. Phys.}\ }\textbf
  {\bibinfo {volume} {87}},\ \bibinfo {pages} {1419} (\bibinfo {year}
  {2015})}\BibitemShut {NoStop}%
\bibitem [{\citenamefont {Christensen}\ \emph {et~al.}(2019)\citenamefont
  {Christensen}, \citenamefont {Wilen}, \citenamefont {Opremcak}, \citenamefont
  {Nelson}, \citenamefont {Schlenker}, \citenamefont {Zimonick}, \citenamefont
  {Faoro}, \citenamefont {Ioffe}, \citenamefont {Rosen}, \citenamefont {DuBois}
  \emph {et~al.}}]{christensen2019}%
  \BibitemOpen
  \bibfield  {author} {\bibinfo {author} {\bibfnamefont {B.}~\bibnamefont
  {Christensen}}, \bibinfo {author} {\bibfnamefont {C.}~\bibnamefont {Wilen}},
  \bibinfo {author} {\bibfnamefont {A.}~\bibnamefont {Opremcak}}, \bibinfo
  {author} {\bibfnamefont {J.}~\bibnamefont {Nelson}}, \bibinfo {author}
  {\bibfnamefont {F.}~\bibnamefont {Schlenker}}, \bibinfo {author}
  {\bibfnamefont {C.}~\bibnamefont {Zimonick}}, \bibinfo {author}
  {\bibfnamefont {L.}~\bibnamefont {Faoro}}, \bibinfo {author} {\bibfnamefont
  {L.}~\bibnamefont {Ioffe}}, \bibinfo {author} {\bibfnamefont
  {Y.}~\bibnamefont {Rosen}}, \bibinfo {author} {\bibfnamefont
  {J.}~\bibnamefont {DuBois}}, \emph {et~al.},\ }\bibfield  {title} {\bibinfo
  {title} {Anomalous charge noise in superconducting qubits},\ }\href@noop {}
  {\bibfield  {journal} {\bibinfo  {journal} {arXiv:1905.13712}\ } (\bibinfo
  {year} {2019})}\BibitemShut {NoStop}%
\bibitem [{\citenamefont {Kuhlmann}\ \emph {et~al.}(2013)\citenamefont
  {Kuhlmann}, \citenamefont {Houel}, \citenamefont {Ludwig}, \citenamefont
  {Greuter}, \citenamefont {Reuter}, \citenamefont {Wieck}, \citenamefont
  {Poggio},\ and\ \citenamefont {Warburton}}]{Kuhlmann2013}%
  \BibitemOpen
  \bibfield  {author} {\bibinfo {author} {\bibfnamefont {A.~V.}\ \bibnamefont
  {Kuhlmann}}, \bibinfo {author} {\bibfnamefont {J.}~\bibnamefont {Houel}},
  \bibinfo {author} {\bibfnamefont {A.}~\bibnamefont {Ludwig}}, \bibinfo
  {author} {\bibfnamefont {L.}~\bibnamefont {Greuter}}, \bibinfo {author}
  {\bibfnamefont {D.}~\bibnamefont {Reuter}}, \bibinfo {author} {\bibfnamefont
  {A.~D.}\ \bibnamefont {Wieck}}, \bibinfo {author} {\bibfnamefont
  {M.}~\bibnamefont {Poggio}},\ and\ \bibinfo {author} {\bibfnamefont {R.~J.}\
  \bibnamefont {Warburton}},\ }\bibfield  {title} {\bibinfo {title} {Charge
  noise and spin noise in a semiconductor quantum device},\ }\href@noop {}
  {\bibfield  {journal} {\bibinfo  {journal} {Nat. Phys.}\ }\textbf {\bibinfo
  {volume} {9}},\ \bibinfo {pages} {570} (\bibinfo {year} {2013})}\BibitemShut
  {NoStop}%
\bibitem [{\citenamefont {Safavi-Naini}\ \emph {et~al.}(2011)\citenamefont
  {Safavi-Naini}, \citenamefont {Rabl}, \citenamefont {Weck},\ and\
  \citenamefont {Sadeghpour}}]{Safavi-Naini2011}%
  \BibitemOpen
  \bibfield  {author} {\bibinfo {author} {\bibfnamefont {A.}~\bibnamefont
  {Safavi-Naini}}, \bibinfo {author} {\bibfnamefont {P.}~\bibnamefont {Rabl}},
  \bibinfo {author} {\bibfnamefont {P.~F.}\ \bibnamefont {Weck}},\ and\
  \bibinfo {author} {\bibfnamefont {H.~R.}\ \bibnamefont {Sadeghpour}},\
  }\bibfield  {title} {\bibinfo {title} {Microscopic model of
  electric-field-noise heating in ion traps},\ }\href@noop {} {\bibfield
  {journal} {\bibinfo  {journal} {Phys. Rev. A}\ }\textbf {\bibinfo {volume}
  {84}},\ \bibinfo {pages} {023412} (\bibinfo {year} {2011})}\BibitemShut
  {NoStop}%
\bibitem [{\citenamefont {Bluvstein}\ \emph {et~al.}(2019)\citenamefont
  {Bluvstein}, \citenamefont {Zhang},\ and\ \citenamefont
  {Jayich}}]{bluvstein2019}%
  \BibitemOpen
  \bibfield  {author} {\bibinfo {author} {\bibfnamefont {D.}~\bibnamefont
  {Bluvstein}}, \bibinfo {author} {\bibfnamefont {Z.}~\bibnamefont {Zhang}},\
  and\ \bibinfo {author} {\bibfnamefont {A.~C.~B.}\ \bibnamefont {Jayich}},\
  }\bibfield  {title} {\bibinfo {title} {Identifying and mitigating charge
  instabilities in shallow diamond nitrogen-vacancy centers},\ }\href@noop {}
  {\bibfield  {journal} {\bibinfo  {journal} {Phys. Rev. Lett.}\ }\textbf
  {\bibinfo {volume} {122}},\ \bibinfo {pages} {076101} (\bibinfo {year}
  {2019})}\BibitemShut {NoStop}%
\end{thebibliography}%


\begin{thebibliography}{0}%
\makeatletter
\providecommand \@ifxundefined [1]{%
 \@ifx{#1\undefined}
}%
\providecommand \@ifnum [1]{%
 \ifnum #1\expandafter \@firstoftwo
 \else \expandafter \@secondoftwo
 \fi
}%
\providecommand \@ifx [1]{%
 \ifx #1\expandafter \@firstoftwo
 \else \expandafter \@secondoftwo
 \fi
}%
\providecommand \natexlab [1]{#1}%
\providecommand \enquote  [1]{``#1''}%
\providecommand \bibnamefont  [1]{#1}%
\providecommand \bibfnamefont [1]{#1}%
\providecommand \citenamefont [1]{#1}%
\providecommand \href@noop [0]{\@secondoftwo}%
\providecommand \href [0]{\begingroup \@sanitize@url \@href}%
\providecommand \@href[1]{\@@startlink{#1}\@@href}%
\providecommand \@@href[1]{\endgroup#1\@@endlink}%
\providecommand \@sanitize@url [0]{\catcode `\\12\catcode `\$12\catcode
  `\&12\catcode `\#12\catcode `\^12\catcode `\_12\catcode `\%12\relax}%
\providecommand \@@startlink[1]{}%
\providecommand \@@endlink[0]{}%
\providecommand \url  [0]{\begingroup\@sanitize@url \@url }%
\providecommand \@url [1]{\endgroup\@href {#1}{\urlprefix }}%
\providecommand \urlprefix  [0]{URL }%
\providecommand \Eprint [0]{\href }%
\providecommand \doibase [0]{http://dx.doi.org/}%
\providecommand \selectlanguage [0]{\@gobble}%
\providecommand \bibinfo  [0]{\@secondoftwo}%
\providecommand \bibfield  [0]{\@secondoftwo}%
\providecommand \translation [1]{[#1]}%
\providecommand \BibitemOpen [0]{}%
\providecommand \bibitemStop [0]{}%
\providecommand \bibitemNoStop [0]{.\EOS\space}%
\providecommand \EOS [0]{\spacefactor3000\relax}%
\providecommand \BibitemShut  [1]{\csname bibitem#1\endcsname}%
\let\auto@bib@innerbib\@empty
\end{thebibliography}%


\providecommand{\noopsort}[1]{}\providecommand{\singleletter}[1]{#1}%
%

\end{document}


\preprint{APS/123-QED}

\title{Supplemental Materials for ``Fast relaxation on qutrit transitions of nitrogen-vacancy centers in nanodiamonds"}

\author{A.~Gardill}
\thanks{These authors contributed equally.}

\author{M.~C.~Cambria}
\thanks{These authors contributed equally.}

\author{S.~Kolkowitz}
\email{kolkowitz@wisc.edu}
 
\affiliation{%
 Department of Physics, University of Wisconsin, Madison, Wisconsin 53706, USA
}

\maketitle

\section{\label{sec:ExperimentalApparatus}Experimental apparatus}
\label{sec:ExpDetails}
Our apparatus consists of a room temperature confocal microscope with a 1.3 NA oil-immersion objective. Two signal generators were used to drive separate state selective \(\pi\text{-pulse}\)s between $\ket{H;0}\leftrightarrow\ket{H;+1}$ and $\ket{H;0}\leftrightarrow\ket{H;-1}$. Software to control the experiment was built upon LabRAD.

\section{\label{sec:Nanodiamond}Silicon Substrate Measurement}

\begin{figure}[h]
\includegraphics[width=0.48\textwidth]{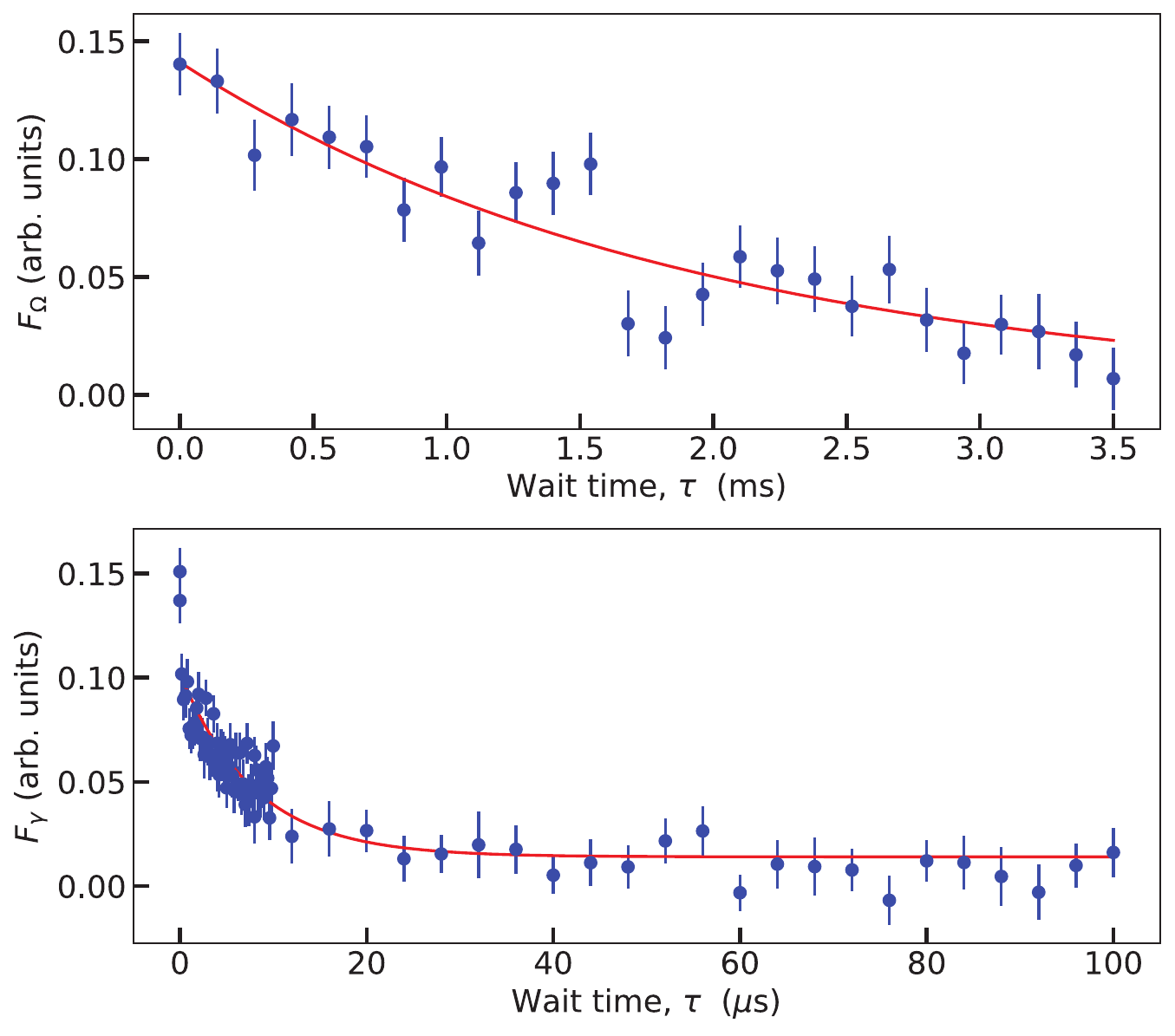}
\caption{Measurement of $\Omega$ (upper panel) and $\gamma$ (lower panel) on a single NV in a nanodiamond on a silicon substrate. Red lines are fits to Eqs.~3 and 4 from the main text, with rates $\Omega = 0.17(3)$ kHz and $\gamma = 63(10)$ kHz. Error bars represent one standard error. \label{fig:silicon}}
\end{figure}

To verify that the noise we observe from all 5 NVs measured is not unique to the glass substrate that hosts the NVs, or the poly-vinyl alcohol (PVA) used in the solution, we measured a single NV deposited on a silicon wafer. This NV was in a nanodiamond solution mixed purely in deionized water containing no PVA. This solution was dropped onto a clean silicon wafer and heated on a hot-plate at $160^\circ$C to evaporate the water. Figure~\ref{fig:silicon} shows a measurement of the relaxation rates of this NV at $\Delta_{\pm}$ = 13.9(6)~MHz, which shows $\gamma$~= 63(10)~kHz and $\Omega$ = 0.17(3)~kHz, confirming that the fast relaxation behavior we observe is intrinsic to the nanodiamonds.

\section{\label{sec:Characterization}NV selection}

\begin{figure}[h]
\includegraphics[width=0.48\textwidth]{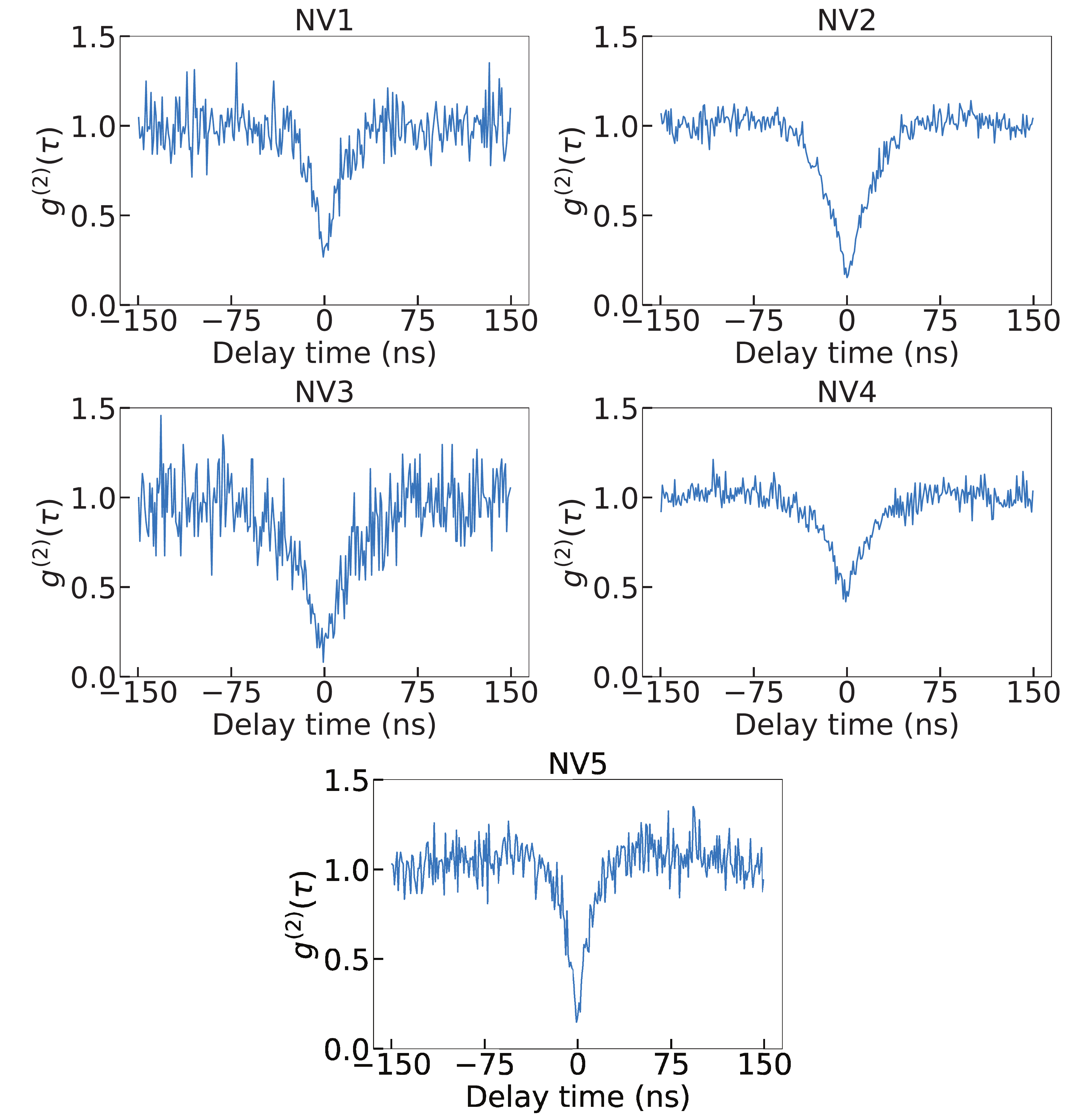}
\caption{\label{fig:g2} Second-order photon correlation function (\(g^{(2)}(\tau)\)) measurements of the five NVs presented in this paper, with no background subtraction. The \(g^{(2)}(0)\) values for the NVs are: NV1: 0.319, NV2: 0.153, NV3: 0.216, NV4: 0.476, NV5: 0.180.}
\end{figure}

Each NV presented in this paper was selected based on its second-order correlation function \(g^{(2)}(\tau)\) and spin-dependent contrast. Figure~\ref{fig:g2} shows the \(g^{(2)}(\tau)\) measurements of all five NVs used in this work. All 5 NVs exhibit a \(g^{(2)}(0) < 0.5\) with no background subtraction, confirming that they are single photon emitters. We then perform optically detected magnetic resonance (ODMR) on the remaining NVs. A majority of the NVs showed little to no sign of ODMR peaks. We found 5 NVs with ODMR contrast strong enough to successfully perform relaxation measurements with. NV5 showed the lowest contrast of $\sim$10\%. This bias in our selection process was necessary to achieve adequate signal-to-noise ratios in our relaxation measurements, and likely selects for larger nanodiamonds and/or NVs further from the surfaces of the nanodiamonds. 

\section{\label{sec:PopDyn}Population Dynamics}

\begin{figure}[h]
\includegraphics[width=0.48\textwidth]{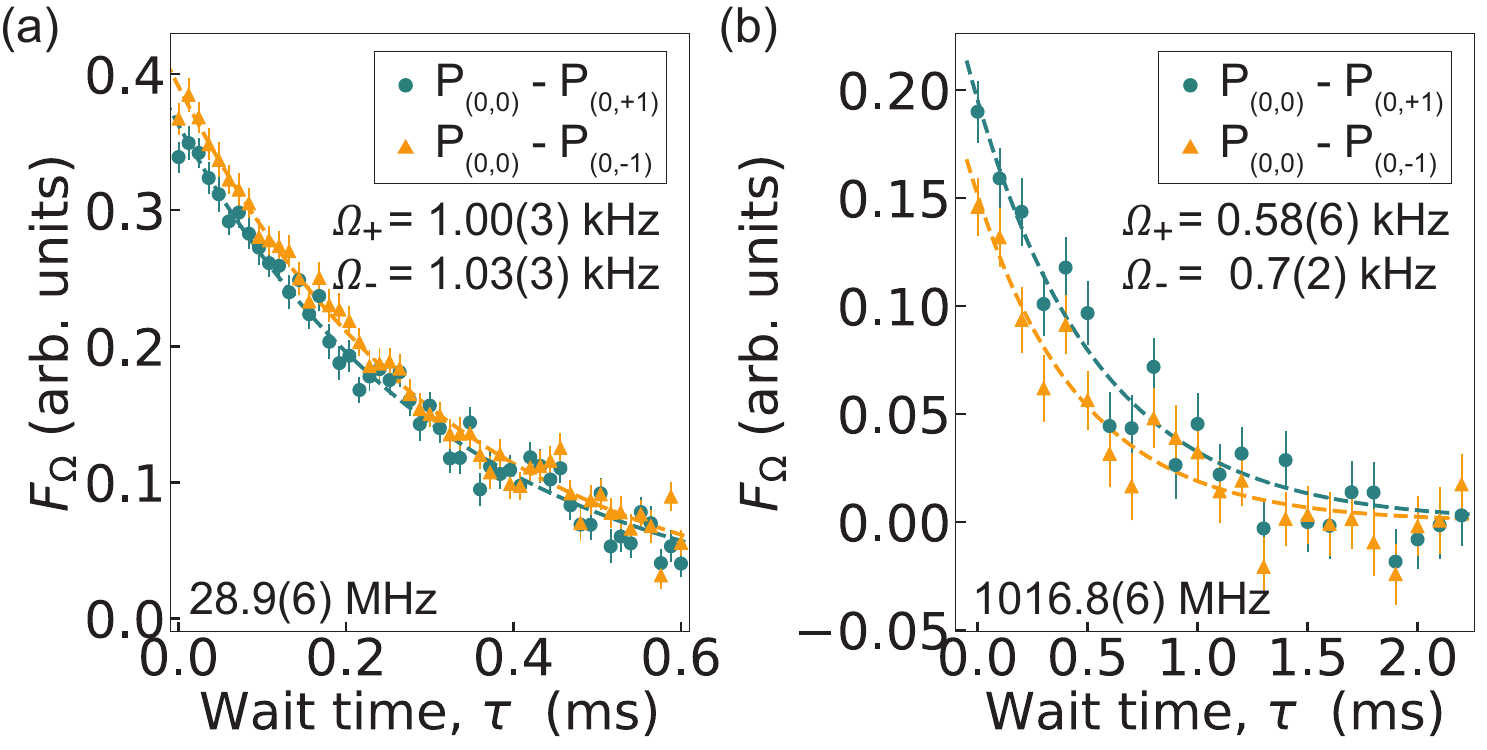}
\caption{\label{fig:omega} Comparison of $\Omega_+$ and $\Omega_-$ for NV1 at two different splittings. Both $P_{0,0} - P_{0,+1}$ (teal circles) and $P_{0,0} - P_{0,-1}$ (orange triangles) are plotted with single exponential fits (dashed lines) to determine $\Omega_+$ and $\Omega_-$, which agree to within error. Error bars represent one standard error. Reported error on $\Delta_\pm$ and $\Omega_\pm$ is twice the standard error. (a) NV1 at $\Delta_\pm=28.9(6)$ MHz. (b) NV1 at $\Delta_\pm=1016.8(6)$ MHz.}
\end{figure}

Here we consider the rate equations for the population dynamics of a generic three level system. There are three possible transitions between states: the $\ket{H;-1} \leftrightarrow \ket{H;+1}$ transition with rate $\gamma$, the $\ket{H;0} \leftrightarrow \ket{H;+1}$ transition with rate $\Omega_+$, and the $\ket{H;0} \leftrightarrow \ket{H;-1}$ transition  with rate $\Omega_-$. Figure~\ref{fig:omega} shows measurements of $\Omega_{+}$ and $\Omega_{-}$ on NV1 at two different splittings, $\Delta_\pm=28.9(6)$ MHz and $\Delta_\pm=1016.8(6)$ MHz. The measured values of $\Omega_{+}$ and $\Omega_{-}$ agree to within error at both splittings, and we take $\Omega_+ = \Omega_- \equiv \Omega$ for all the splittings considered in this paper ($\sim10-1000$ MHz). This assumption was consistent with all of our measurements on all of the NVs in this work. The reported values of $\Omega$ are measured on the  $\ket{H;0} \leftrightarrow \ket{H;+1}$ transition. The system of equations describing the change of population $\rho_i$ in state $\ket{H;i}$ is
\begin{equation}\label{pop_matrix}
\frac{d}{dt}
\begin{pmatrix} 
\rho_0 \\
\rho_{+1} \\
\rho_{-1}
\end{pmatrix} =
\begin{pmatrix} 
-2\Omega & \Omega & \Omega \\
\Omega & -\Omega - \gamma  & \gamma \\
\Omega &  \gamma & -\Omega - \gamma
\end{pmatrix}
\begin{pmatrix} 
\rho_0 \\
\rho_{+1} \\
\rho_{-1}
\end{pmatrix}.
\end{equation}
Requiring that $\rho_0(\tau) + \rho_{+1}(\tau) + \rho_{-1}(\tau) = 1$ and setting the initial condition $\rho_i(0)$  results in the population dynamics given by Eqs.~1 and 2 in the main text.

\section{\label{sec:9Measurements}Complete set of relaxation measurements}
\begin{figure}[h]
\includegraphics[width=0.48\textwidth]{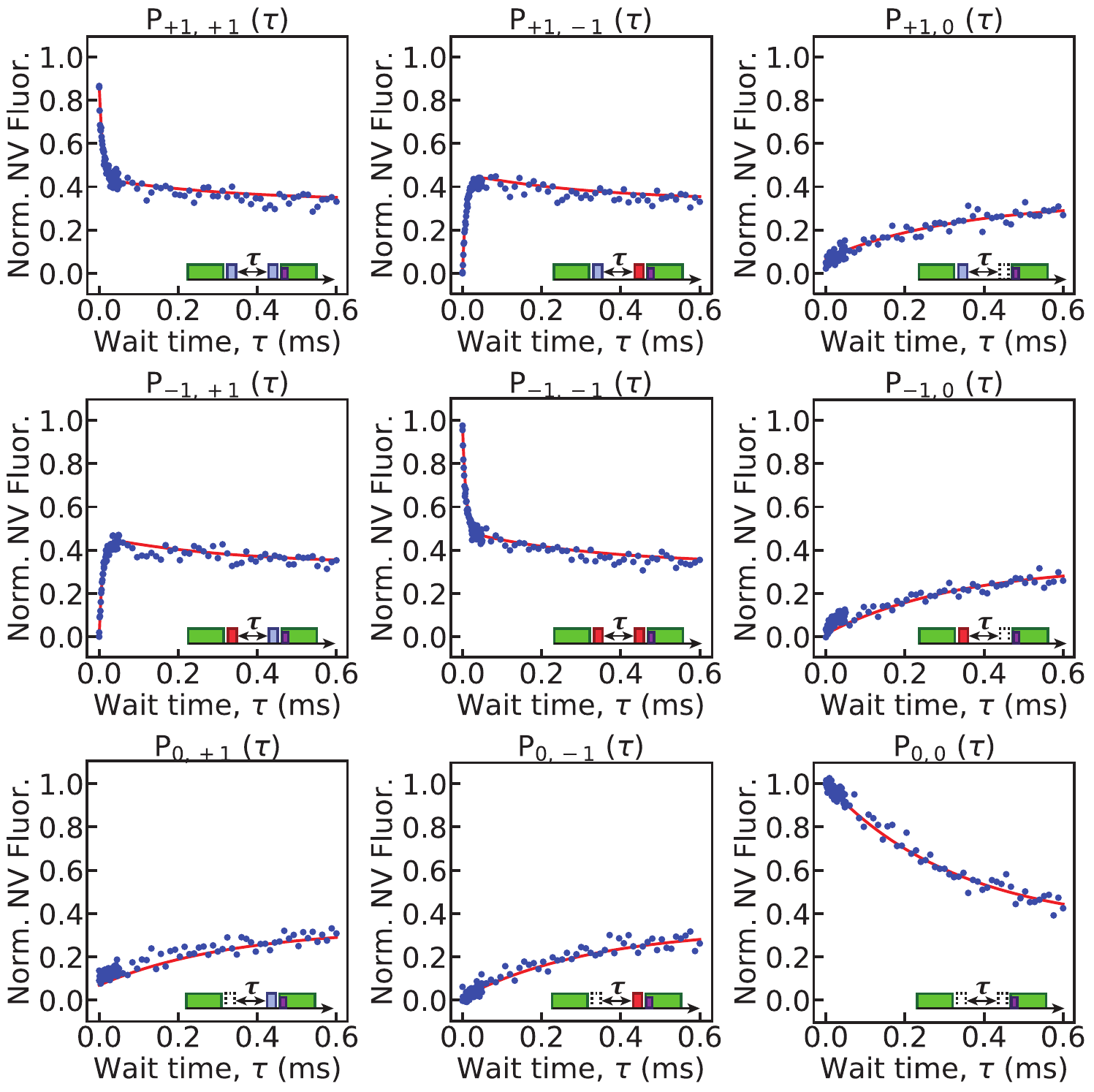}
\caption{\label{fig:9measurements} Example of all 9 measurements made possible by the ability to prepare and readout in any of the three NV states. These measurements were conducted at $\Delta_{\pm}$ = 28.9(6) MHz. Red lines are Eqs.~\ref{pm_infidelity} and \ref{zero_infidelity}, with $\epsilon_+ = 6.9\%$,  $\epsilon_- = 1.3\%$, $\Omega = 1.0$ kHz, and $\gamma$ = 56 kHz.}
\end{figure}

The ability to prepare and readout in any of the three NV spin states allows for a total of nine possible relaxation measurements. To confirm our model, we performed all nine possible measurements on the same NV with the same applied magnetic field, as shown in Fig.~\ref{fig:9measurements}. The population dynamics for all 9 possible combinations are well described by the population dynamics given by Eqs.~1 and 2 in the main text after accounting for $\pi$-pulse infidelities, as discussed below in Sec.~\ref{sec:Infidelity}. We denote the normalized fluorescence of a relaxation measurement with initialization into $\ket{H;i}$ and readout of $\ket{H;j}$ by $P_{i,j}$. The data was normalized such that zero corresponds to $P_{+1,-1}(\tau=0)$ and unity corresponds to $P_{0,0}(\tau=0)$.


\section{\label{sec:SubMethod} Protocol for extracting the rates $\gamma$ and $\Omega$}

\setlength{\tabcolsep}{9pt}
\begin{table*}[b]
\caption{\label{tab:supp_rel_rates}
Complete set of relaxation rates presented in this work. Reported uncertainty is twice the standard error. The angle \(\theta_{B}\) is the estimated angle of the applied magnetic field with respect to the NV axis. This value is calculated numerically from the NV resonances as the applied magnetic field is increased with a fixed orientation. Measurements taken in absence of applied magnetic field are not assigned \(\theta_{B}\). Measurements marked with an asterisk (*) or dagger (\(\dagger\)) are used as evidence against magnetic noise driving transitions between $\ket{H;\pm1}$, as discussed in the main text.
}
\begin{ruledtabular}
\begin{tabular}{cccc|cccc}
\multicolumn{4}{c|}{~~~~NV1}&\multicolumn{4}{c}{~~~~NV2}\\
\textrm{\(\Delta_{\pm}\) (MHz)}&
\textrm{\(\Omega\) (kHz)}&
\textrm{\(\gamma\) (kHz)}&
\textrm{\(\theta_B\) (deg)}&
\textrm{\(\Delta_{\pm}\) (MHz)}&
\textrm{\(\Omega\) (kHz)}&
\textrm{\(\gamma\) (kHz)}&
\textrm{\(\theta_B\) (deg)}\\

\hline

19.5(6) & 0.83(8) & 58(3) & n/a &
15.3(6) & 0.24(2) & 124(6) & n/a \\

19.8(6) & 1.28(8) & 117(8) & 37 &
29.1(6) & 0.41(2) & 20.9(6) & 32 \\

27.7(6) & 1.30(12) & 65(3) & 37 &
29.2(6) & 0.33(3) & 31.1(8) & 64 \\

28.9(6) & 1.00(3) & 56(3) & 37 &
44.8(6) & 0.36(2) & 6.4(2) & 32 \\

32.7(6) & 1.42(10) & 42.6(18) & 37 &
45.5(6) & 0.27(2) & 8.5(2) & 64 \\

51.8(6) & 1.85(16) & 13.1(4) & 37 &
56.2(6) & 0.326(16) & 3.64(16) & 32 \\

97.8(6) & 1.41(10) & 3.9(2) & 37 &
56.9(6) & 0.42(10) & 3.77(18) & 32 \\

116.0(6) & 1.18(12) & 4.7(2) & 37 &
85.2(6) & 0.29(2) & 2.62(10) & 64 \\

268.0(6) & 1.04(8) & 2.0(2) & 37 &
101.6(6) & 0.312(18) & 1.33(10) & 32 \\

350.0(6)$\dagger$& 0.72(8) & 1.6(2) & 58 &
280.4(6) & 0.28(2) & 0.44(3) & 64 \\

561.7(6) & 1.19(12) & 0.70(10) & 37 &
697.5(6) & 0.29(4) & 0.81(12) & 64 \\

1016.8(6)*& 0.58(6) & 0.41(10) & 37 &
\multicolumn{4}{c}{} \\

\hline

\multicolumn{4}{c|}{~~~~NV3}&\multicolumn{4}{c}{~~~~NV4}\\
\textrm{\(\Delta_{\pm}\) (MHz)}&
\textrm{\(\Omega\) (kHz)}&
\textrm{\(\gamma\) (kHz)}&
\textrm{\(\theta_B\) (deg)}&
\textrm{\(\Delta_{\pm}\) (MHz)}&
\textrm{\(\Omega\) (kHz)}&
\textrm{\(\gamma\) (kHz)}&
\textrm{\(\theta_B\) (deg)}\\

\hline

17.1(6) & 0.7(3) & 110(20) & 51 &
23.4(1) & 0.28(3) & 35(3) & n/a \\

28.6(6) & 0.53(10) & 90(10) & 51 &
26.2(6) & 0.33(6) & 29(2) & 9 \\

53.0(6) & 0.87(18) & 26.2(18) & 51 &
36.2(6) & 0.32(6) & 20.3(10) & 9 \\

81.2(6) & 1.7(4) & 17.5(12) & 51 &
60.5(6) & 0.24(4) & 9.1(6) & 9 \\

128.0(6) & 0.60(10) & 11.3(8) & 51 &
48.1(6) & 0.31(2) & 15.8(6) & 9 \\

283.1(6) & 0.70(14) & 5.6(6) & 51 &
92.3(6) & 0.25(2) & 6.4(2) & 9 \\

495.8(6) & 1.4(8) & 3.7(8) & 51 &
150.8(1) & 0.29(4) & 4.1(3) & 9 \\

746.0(6) & 1.0(3) & 2.8(6) & 51 &
329.6(6) & 0.33(4) & 1.23(14) & 9 \\

\multicolumn{4}{c|}{} &
884.9(6) & 0.29(4) & 0.45(6) & 9 \\

\multicolumn{4}{c|}{} &
1080.5(6) & 0.28(10) & 0.7(2) & 9 \\

\multicolumn{4}{c|}{} &
1148.4(6) & 0.38(8) & 0.35(6) & 9 \\

\hline

\multicolumn{4}{c|}{~~~~NV5}&\multicolumn{4}{c}{}  \\
\textrm{\(\Delta_{\pm}\) (MHz)}&
\textrm{\(\Omega\) (kHz)}&
\textrm{\(\gamma\) (kHz)}&
\textrm{\(\theta_B\) (deg)}\\
\cline{1-4}

10.9(6) & 0.45(12) & 240(50) & 71 &
\multicolumn{4}{c}{} \\

23.1(6) & 1.0(3) & 62(16) & 71 &
\multicolumn{4}{c}{} \\

29.8(6) & 1.01(18) & 19(2) & 71 &
\multicolumn{4}{c}{} \\

51.9(6) & 0.39(8) & 18(3) & 71 &
\multicolumn{4}{c}{} \\

72.4(6) & 0.8(2) & 16(2) & 71 &
\multicolumn{4}{c}{} \\

112.9(6) & 0.9(3) & 12.1(18) & 71 &
\multicolumn{4}{c}{} \\

164.1(6) & 0.7(2) & 5.6(10) & 71 &
\multicolumn{4}{c}{} \\

256.2(6) & 	0.23(8) & 2.1(6) & 71 &
\multicolumn{4}{c}{} \\

\end{tabular}
\end{ruledtabular}
\end{table*}

\begin{figure}[h]
\includegraphics[width=0.48\textwidth]{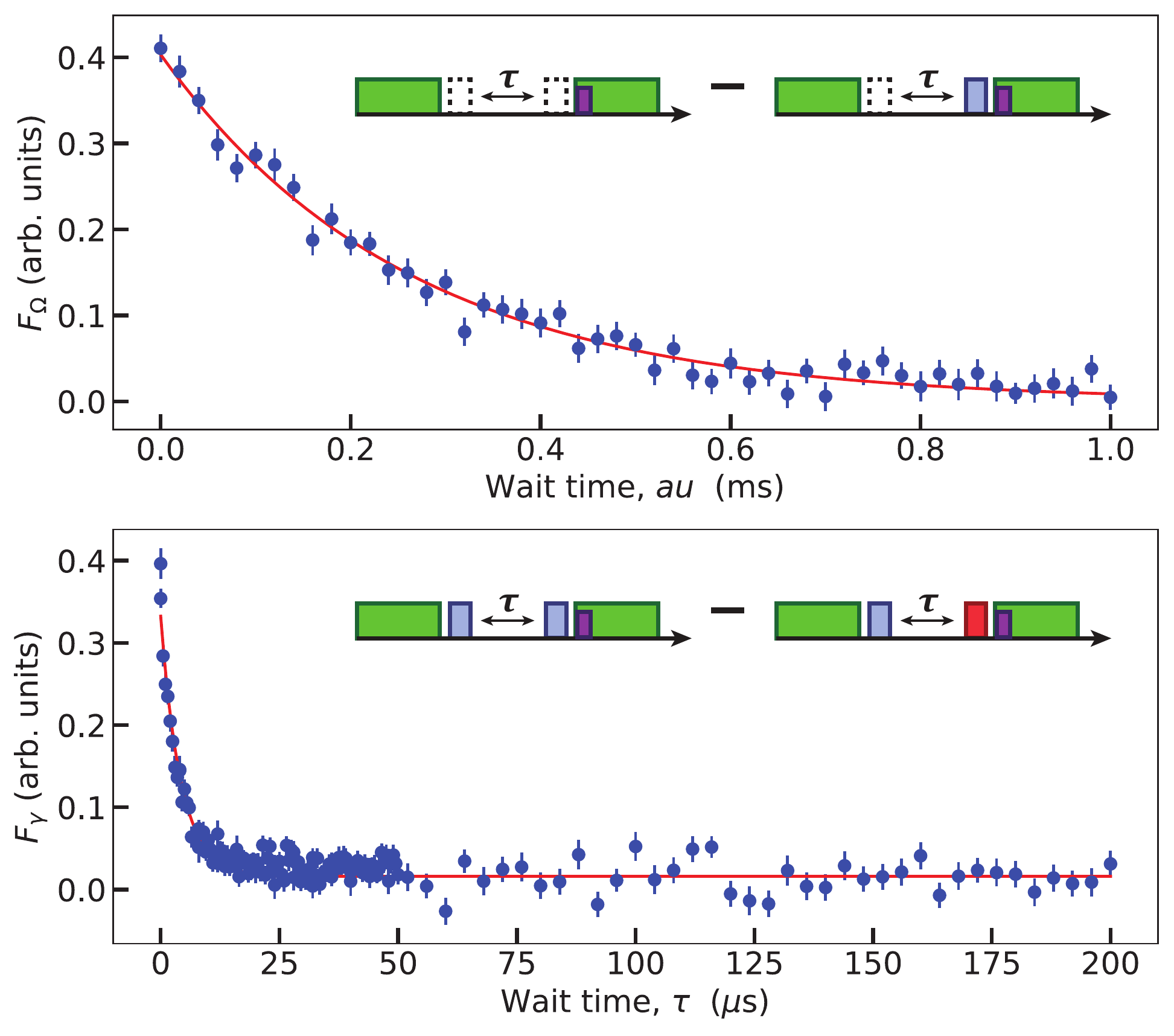}
\caption{Representative data from NV1 showing subtraction of relaxation signals to extract \(\Omega\) (upper panel) and \(\gamma\) (lower panel), where red lines are fits to Eqs.~3 and 4 from the main text. The splitting is $\Delta_\pm=19.8(6) \ \text{MHz}$ and the rates extracted are \(\Omega\) = 1.28(8) and \(\gamma\) = 117(8) kHz. Error bars are one standard error.}
\label{fig:subtraction}
\end{figure}

To extract $\gamma$ and $\Omega$, we take the difference between measured population decay curves and fit to the result with the single exponential functions $F_\Omega$ and $F_\gamma$ defined in Eqs.~3 and 4 of the main text. At each value of $\Delta_{\pm},$ we measure $P_{+1,+1}$, $P_{+1,-1}$, $P_{0,+1}$, and $P_{0,0}$, with $P_{+1,+1}$ and $P_{+1,-1}$ measured on two timescales in order to sufficiently resolve the fast \(\ket{H;-1} \leftrightarrow \ket{H;+1}\) relaxation. Figure~\ref{fig:subtraction} shows representative data for $F_\Omega$ and $F_\gamma$. Table~\ref{tab:supp_rel_rates} shows all the rates measured for this paper. The measurements for all 5 NVs were recorded in pseudo-random order to avoid conflating temporal effects with dependence on \(\Delta_{\pm}\). 

\section{\label{sec:Infidelity} Microwave \(\pi\)-pulse infidelities}

In order to fit the population decay curves shown in Figs.~1 and 2 of the main text and Fig.~\ref{fig:9measurements} of the supplement, imperfections in the microwave $\pi$-pulses used to prepare and readout the populations must be taken into account. The fraction of population not transferred by the \(\pi_{\pm}\text{-pulse}\) is denoted $\epsilon_{\pm}$. The modified equations for reading out the population of $\ket{H;+1}$, $\ket{H;-1}$, and $\ket{H;0}$ after initializing in $\ket{H;+1}$ are

\begin{align}\label{pm_infidelity}
P'_{+1,\pm1}(\tau) &= \frac{1}{3}\pm \left[ \frac{1}{2}(1-\epsilon_+)e^{-(2\gamma+\Omega)\tau}\right](1-\epsilon_\pm)  \\
  &\quad- \left[\frac{1}{2}(\epsilon_+-\frac{1}{3})e^{-3\Omega\tau}\right](1-\epsilon_\pm)\nonumber\\ 
 &\quad+ \left[(\epsilon_+-\frac{1}{3})e^{-3\Omega\tau} \right]\epsilon_\pm ,\nonumber\\
P'_{+1,0}(\tau) &= 
\frac{1}{3} + (\epsilon_+-\frac{1}{3})e^{-3\Omega\tau}~.\label{zero_infidelity} 
\end{align}
These equations are used to produce the dashed colored lines in Figs.~1 and 2 of the main text, with $\epsilon_+ = 6.9\%$ and $\epsilon_- = 1.3\%$. The infidelities were calculated as the relative decrease in contrast in the measured Rabi signal over one \(\pi\text{-pulse}\).

If we include the effect of $\pi\text{-pulse}$ infidelities on $F_\Omega$ and $F_\gamma$, then we obtain the modified equations

\begin{align}
    F_\Omega' &= \left(1-\epsilon_+\right)e^{-3\Omega\tau},
\label{f1_prime}\\
    F_\gamma' &= \left(1-\frac{1}{2}(\epsilon_+ + \epsilon_-)\right) \left(1-\epsilon_+\right)e^{-(2\gamma+\Omega)\tau} \\
    &\quad + \frac{3}{2}\left(\epsilon_+ - \epsilon_-\right)\left(\epsilon_+ - \frac{1}{3}\right)e^{-3\Omega\tau}.\nonumber
\label{f2_prime}
\end{align}
The differential protocol described in Sec.~\ref{sec:SubMethod} for the extraction of the rates $\gamma$ and $\Omega$ is relatively insensitive to $\pi$-pulse infidelities, so they were not included in the analysis in order to simplify the procedure. However, if the difference between the infidelities \(\epsilon_+\) and \(\epsilon_-\) is significant, then \(F_\gamma'\) will not decay to zero on the \((2\gamma+\Omega)\) timescale, resulting in an apparent offset. This was observed in some of our measurements where \(\gamma \gg \Omega\). In these cases we account for the discrepancy by adding a fixed offset to $F_\gamma$. This offset is not a free parameter of the fit, but is instead calculated as the average difference in the subtracted data after $\sim3\times$ the time constant $1/(2\gamma+\Omega)$.

\begin{figure}[b]
\includegraphics[width=0.48\textwidth]{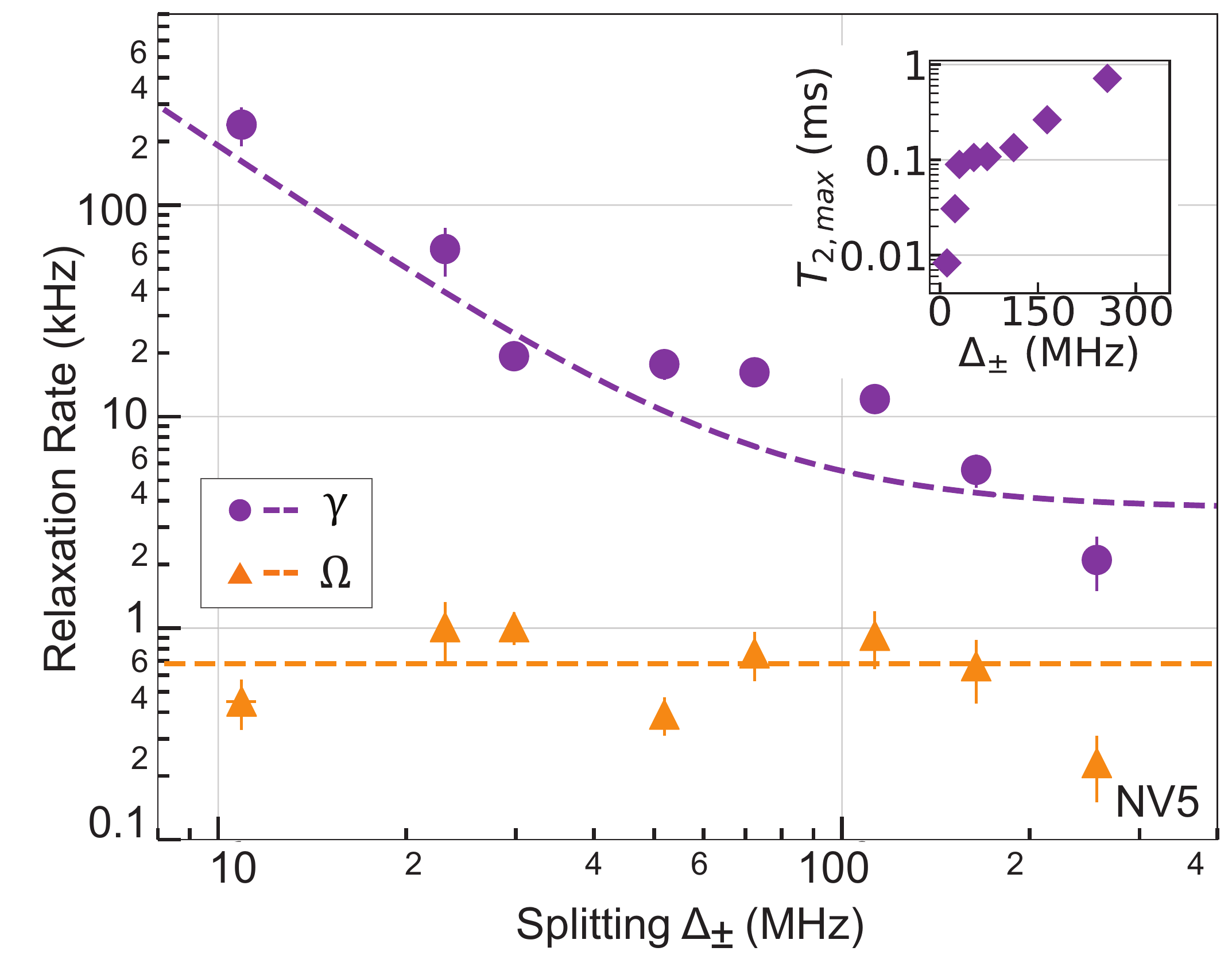}
\caption{Dependence of the relaxation rates of NV5 on the frequency splitting between \(\ket{H;\pm1}\) states, \(\Delta_\pm\). Purple circles represent measurement of \(\gamma\) and orange triangles represent measurement of \(\Omega\). A fit of \(\gamma(\Delta_{\pm}) = A_0/\Delta_{\pm}^{2} + \gamma_\infty\) is shown as purple dashed line. Inset shows the corresponding maximum achievable \(T_\text{2,max}\) (purple diamonds) on a semi-log plot. Error bars are twice the standard error. 
\label{fig:NV5} }
\end{figure}

\section{\label{sec:Error}Error analysis}

An individual data point in a relaxation measurement \(P_{i,j}\) is the average of \(\sim10-100\) runs each consisting of ${\sim10^5}$ repetitions of the same measurement sequence. Each run produces a single value which is the cumulative total of the counts recorded over the measurement sequence repetitions. The standard error of each data point is taken with respect to the runs. The fits to $F_\Omega$ and $F_\gamma$ are weighted by the standard errors on the data points, and the errors on $\Omega$ and $\gamma$ are propagated from the standard errors of the fits to $F_\Omega$ and $F_\gamma$. 


\section{\label{sec:NV5}Data for NV5}

Figure~\ref{fig:NV5} shows the measured values of $\gamma$ and $\Omega$ for NV5 (which were not plotted in the main text due to space constraints). This NV had the lowest spin-dependent fluorescence contrast ($\sim10\%$) which, in combination with imperfect alignment of the applied magnetic field, prohibited us from making measurements at splittings greater than \(\sim240\) MHz. 

\section{\label{sec:LaserVary}Charge Noise Dependence on Laser Power}

\begin{figure}[h]
\includegraphics[width=0.48\textwidth]{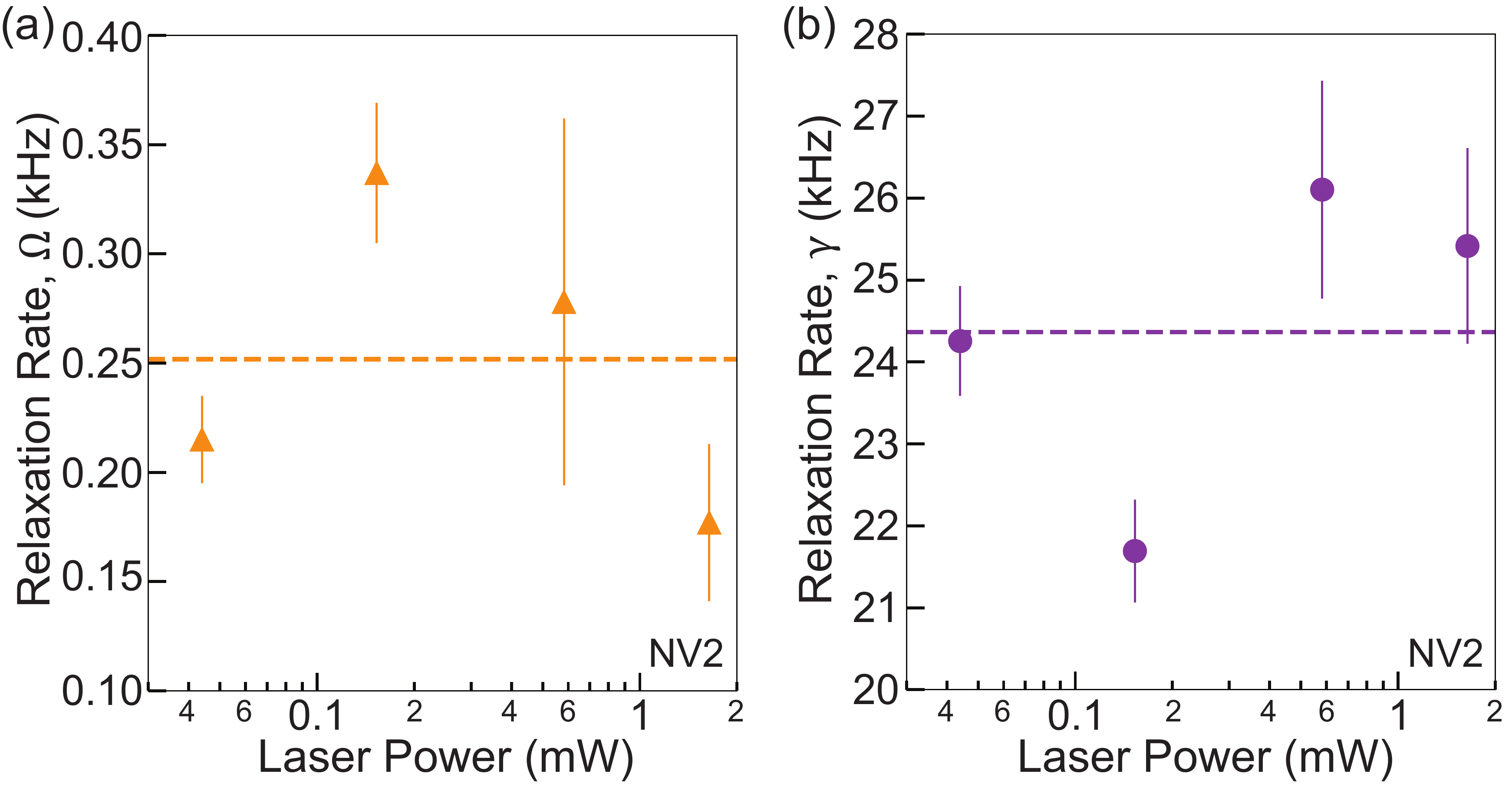}
\caption{Dependence of the relaxation rates on 532 nm laser power. Relaxation rates \(\Omega\) (a) and \(\gamma\) (b) measured on NV2 at four different laser intensities at a splitting of \(\sim30\) MHz. No clear dependence on laser power is observed. Dotted lines represent the average value of these four rates, respectively. Error bars are twice the standard error. 
\label{fig:laser} }
\end{figure}

The power of the 532 nm laser used for polarization and readout was $\sim$ 0.5 mW for all the measurements reported in this paper (aside from the measurements described in this section and shown in Fig.~\ref{fig:laser}). 
While the relaxation we observe occurs in the dark period between the laser pulses, the noise we attribute to the relaxation could be influenced by heating or other transient effects induced by optical excitation. 
To determine the effect of the laser power on the relaxation rates, we performed four measurements on NV2 at a splitting of $\sim31$ MHz for four laser powers spanning a range from $\sim50~\mu$W$-1.5$~mW. Figure~\ref{fig:laser} shows the measured values of $\gamma$ and $\Omega$, with dotted lines representing the average value of all four measurements. The laser power was measured directly before the microscope objective. The variation of $\gamma$ in Fig.~\ref{fig:laser} is similar in magnitude to that variation seen in the long-scale temporal fluctuation measurements at a fixed laser power (shown in Fig.~4 of the main text). We see no scaling of the rates with laser power beyond the expected fluctuations of the rates with time. 
\\
\section{\label{sec:TimeVary}Temporal fluctuations of \(\gamma\) for NV1}

\begin{figure}[h]
\includegraphics[width=0.5\textwidth]{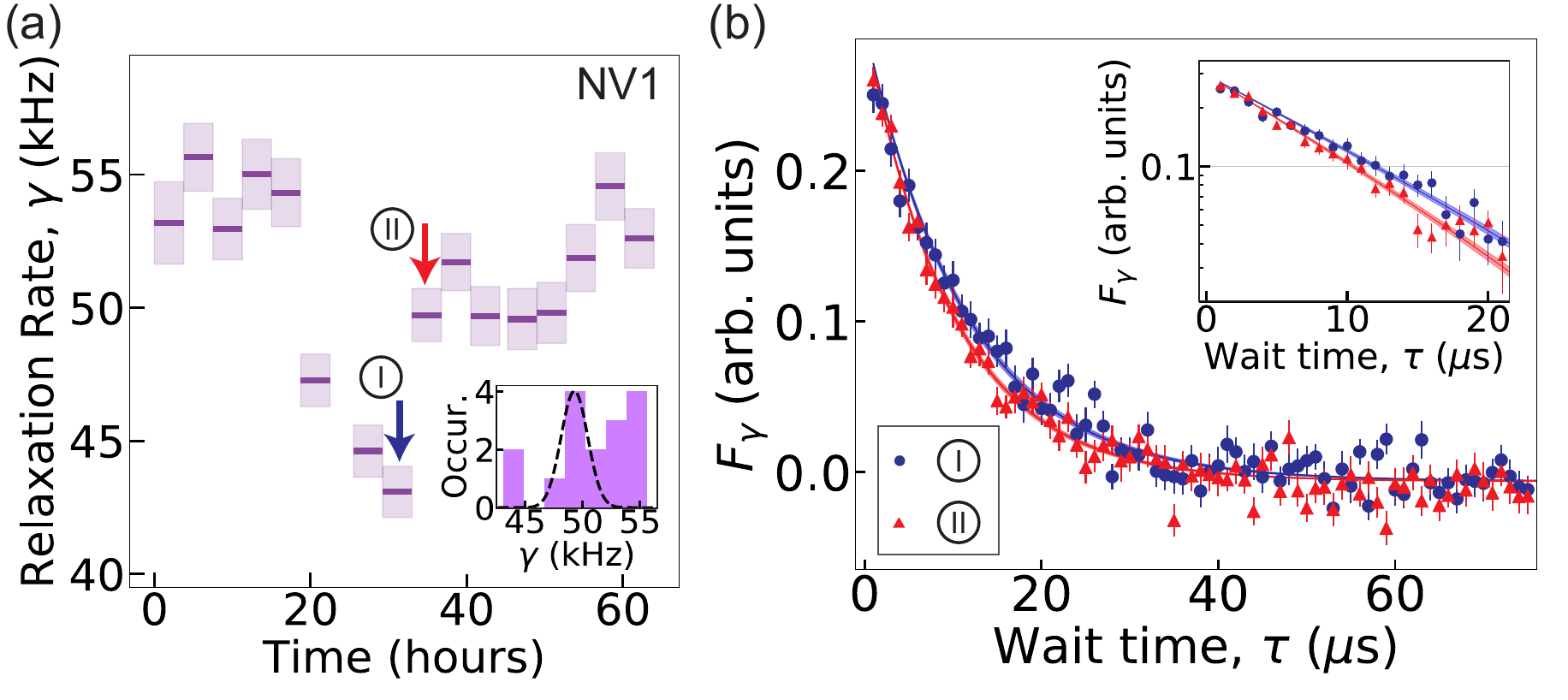}
\caption{Temporal fluctuations in \(\gamma\) observed in NV1. (a) Rate \(\gamma\) at a single splitting \(\Delta_{\pm} \approx 26\) MHz measured over 65 hours. Error bars are one standard error. Inset shows histogram of measured rates overlaid with a Gaussian curve with standard deviation equal to the average standard error of the measurements. (b) Subtraction curves \(F_\gamma\) for adjacent measurements in (a). Inset shows first 20 \(\mu\)s of same data plotted on a semi-log scale. Error bars are one standard error, shaded error on fit represents one standard error of the extracted $\gamma$.  }
\label{fig:time_rates}
\end{figure}

Figure~4 in the main text presents temporal fluctuations of the rate $\gamma$ for NV2 at \(\Delta_{\pm}\approx 29 \ \text{MHz}\). We performed the same measurement on NV1 at  \(\Delta_{\pm}\approx 26 \ \text{MHz}\) and observed similar fluctuations. Figure~\ref{fig:time_rates}(a) shows consecutive measurements performed on NV1 at the same splitting over $\sim65$ hours. The inset histogram shows the distribution of the measured rates along with a Gaussian curve of standard deviation equal to the average standard error of a single measurement. It is clear that the spread of the measured rates is larger than the average uncertainty of the measurements. Figure~\ref{fig:time_rates}(b) compares the population decay curves for two adjacent measured rates, showing that the observed fluctuations in $\gamma$ are above our signal-to-noise ratio.

\bibliographystyle{apsrev4-1}
\bibliography{SuppReferences}